\documentclass{aa}

\usepackage{graphics,epsfig,amsmath,amssymb,amstext,txfonts}

\begin{document}

%\title{Propagation of high-energy cosmic ray nuclei in extragalactic turbulent magnetic fields: energy spectrum and composition}

\title{Constraints on the origin of ultra-high-energy cosmic rays from cosmogenic neutrinos and photons}

\author{Guillaume Decerprit$^1$ and Denis Allard$^2$}

\institute{$^1$DESY, Platanenallee 6, 15738 Zeuthen, Germany\\$^2$Laboratoire Astroparticule et Cosmologie (APC), Universit\'e Paris 7/CNRS, 10 rue A. Domon et L. Duquet, 75205 Paris Cedex 13, France.}

\offprints{guillaume.decerprit@desy.de\\denis.allard@apc.univ-paris7.fr}

\date{}
%\doi{10.1051/0004-6361/201117673}

\abstract{We study the production of cosmogenic neutrinos and photons during the extragalactic propagation of ultra-high-energy cosmic rays (UHECRs). For a wide range of models in cosmological evolution of source luminosity, composition and maximum energy we calculate the expected flux of cosmogenic secondaries by normalizing our cosmic ray output to experimental spectra {and comparing the diffuse flux of GeV-TeV gamma-rays to the experimental one measured by the Fermi satellite}. Most of these models yield significant neutrino fluxes for current experiments like IceCube or Pierre Auger. Furthermore, we discuss the possibilities of signing the presence of UHE proton sources either within or outside the cosmic ray horizon using neutrinos or photons observations even if the cosmic ray composition becomes heavier at the highest energies. We discuss the possible constraints that could be brought on the UHECR origin from the different messengers and energy ranges. 
%\keywords{Cosmic rays; composition; abundances; propagation; neutrinos; photons}
}

\authorrunning{G. Decerprit, D. Allard}
\titlerunning{Constraints on the origin of UHE Cosmic Rays using cosmogenic neutrinos and photons}

\maketitle

%**************************************************************************	
%**************************************************************************
\section{Introduction}
\label{Introduction}

The origin of ultra-high-energy cosmic rays is one of the main topics in high-energy astrophysics. Their origin still remains unknown after decades of experimental efforts. Recent experiments such as the High Resolution fly's eye (HiRes, Abbasi et al., 2004), the Akeno Giant Air Shower Array (AGASA, Nagano et al., 1992) and primarily the Pierre Auger Observatory  (Abraham et al., 2004) have increased the available cosmic ray statistics above $10^{18}$ eV, allowing the first solid studies from the point of view of the spectrum, the composition and the arrival directions of ultra-high-energy cosmic rays (UHECRs). A sharp decrease of the UHECR flux above 3-4$\times 10^{19}$ eV possibly related to the theoretically expected GZK cut-off (Greisen, 1966; Zatsepin \& Kuzmin, 1966) seems to be firmly established, whereas the interpretation of composition or arrival direction data as well as the consistency of the results reported by the different experiments are matters of intense debates. 

A multimessenger approach of the UHECR question is a promising way for solving the mystery of their origin. Indeed, shortly after the prediction of a cut-off at the highest energies caused by the interactions of UHECR protons and nuclei with the cosmic microwave background (CMB) photons, the unavoidable associated production of UHE neutrinos through the decay of charged pions was pointed out by Berezinsky and Zatsepin (1969). Because these {\it cosmogenic neutrinos} can travel from their production site without undergoing interactions or deflection, they were, soon after this pioneering work, considered as potentially interesting probes of the highest energy phenomena in the universe. Their expected flux on earth was intensively calculated in the past decades for different astrophysical scenarios on the cosmological evolution of the sources, the composition or the maximum energy at the sources (see for instance Stecker ,1979; Hill \& Schramm,1985; Engel et al., 2001; Kalashev et al., 2002; Secker \& Stanev, 2005; Hooper et al., 2005; Ave et al., 2005; Stanev et al., 2006; Allard et al., 2006; Anchordoqui et al., 2007; Takami et al., 2009; Berezinsky, 2009; Ahlers et al., 2009; Kotera et al., 2010).

Cosmogenic gamma-rays are also produced during the propagation of UHECRs. Unlike neutrinos, these very high-energy gamma-rays interact rapidly and produce electromagnetic cascades. As a result the universe is opaque to gamma-rays from a few hundreds of GeV to a few $10^{18}$ eV. Above $10^{19}$ eV the universe becomes more and more transparent to photons and very high-energy gamma-rays can propagate a few tens of megaparsecs without losing a great amount of energy. As a result, these very high-energy cosmogenic gamma-rays were discussed in the literature either as signatures of the so-called Top-Down models (see for instance  Protheroe \& Johnson, 1996; Lee, 1998; Sigl et al., 1999; Semikoz \& Sigl., 2004) or as probes of UHECR acceleration in the local universe (e.g, Yoshida \& Teshima, 1993; Protheroe and Johnson, 1996; Lee, 1998; Semikoz \& Sigl., 2004; Gelmini et al., 2007a and 2007b, Taylor \& Aharonian, 2009; Taylor et al., 2009; Kuempel et al., 2009; Hooper et al., 2011; Ahlers \& Salvado, 2011).

Because electromagnetic cascades, piling up below 100 GeV, are produced during UHECR propagation and are likely to be associated with the production of cosmogenic neutrinos, it has soon been realized that measurements of the diffuse gamma-ray background could allow one to put constraints on the cosmological evolution of the UHECR luminosity (Strong et al., 1973, 1974) and on the maximum allowable cosmogenic neutrino fluxes (Berezinsky \& Smirnov, 1975). Modern versions of this calculation were attempted using the EGRET measurements (Sreekumar et al., 1998; Strong et al., 2004)  in Kalashev et al., (2002), Semikoz \& Sigl (2004), Kalashev et al. (2007). More recently, the Fermi satellite measurements  (Abdo et al., 2010) reported a gamma-ray background between 100 MeV and 100 GeV lower than previously estimated using EGRET data. The additional constraints allowed by this new measurement were discussed by Berezinsky et al. (2010), Ahlers et al. (2010) and more recently by Wang et al. (2011). These studies agree that the constraints implied by this new estimate of the gamma-ray background are much more stringent than those using EGRET data, but their conclusions differ on the impact of these new constraints on the observability of UHE neutrino fluxes (see below). 
In this paper we study the production of cosmogenic photons and neutrinos assuming different astrophysical scenarios for the cosmological evolution of the source luminosity, the composition and different scenarios for the transition from galactic to extragalactic cosmic rays in the same spirit as in Kotera et al. (2010). We discuss the constraints brought by the Fermi measurements of the diffuse extragalactic background on the expectations for cosmogenic neutrinos or UHE photons detection. Furthermore, we discuss the possible neutrino or photon signatures from UHECR proton accelerators either in the local universe or at cosmological distances (diffuse or point sources) even if the UHECR composition becomes heavier, which is one of the plausible interpretations of the recent measurement of the longitudinal development of air showers (hereafter $X_{max}$ measurements) of the Pierre Auger Observatory (Abraham et al. 2010a). In the next section, we briefly introduce our code that simulates the propagation of UHECR protons and nuclei and our Monte Carlo tool for the development of intergalactic electromagnetic cascades. In section 3 we then calculate cosmogenic photons and neutrino fluxes by normalizing our simulations with the experimental spectrum measured by the Pierre Auger Observatory (Abraham et al., 2010b).  We finally discuss these results and the possible constraints on the UHECR origin that can potentially be achieved by observing either point sources or diffuse contributions of cosmogenic secondaries.

%\begin{figure*}[ht]
%\centering{\includegraphics[width=\textwidth]{8653fig1.pdf}}
%\caption{Structure of a purely turbulent magnetic field with variance $\sqrt{\langle \|\mathbf{\vec{B}^2}\| \rangle} = 10$~nG and maximum turbulence scale $\lambda_{\mathrm{max}} = 1$~Mpc. From left to right: distribution of the magnetic field squared norm, $\|\mathbf{\vec{B}^2}\|$, distribution of component $B_x$, and magnetic field two-point autocorrelation as a function of point separation.}
%\label{fig:fieldStructure}
%\end{figure*}

\section{Modeling the propagation of cosmic rays and electromagnetic cascades} 

In the following we consider the propagation of UHECR proton and nuclei in the intergalactic medium, the production of secondary photons, pairs and neutrinos and the development of electromagnetic cascade within our Monte Carlo framework. In addition to the CMB photons, we consider in these calculations all other main photon targets in the intergalactic medium. The universal radio background (URB) is considered in this study only for the development of electromagnetic cascades; for this purpose we conservatively use the spectrum given in Clark et al. (1970) and the cosmological evolution estimated in Lee (1998). Several authors have already treated the impact of different estimates of the radio background on the output of electromagnetic cascades (see for instance Gelmini et al., 2008; Taylor and Aharonian, 2009). None of the conclusions of our studies should be significantly affected by the uncertainties related to the URB.  In the infra-red, optical and ultra-violet range (hereafter IR/Opt/UV) we use the model of the density and cosmological evolution of the background photons published by Kneiske et al. (2004). We use the latest available update of their best-fit model, which proved to be compatible with {the Fermi} observations of distant blazars and GRBs (Abdo et al., 2010). 
For the sake of comparison, we also  refer to the estimate of Stecker et al. (2006) used in our previous works. This estimate was, however, challenged by  {the recent Fermi }observations suggesting a fainter photon density in optical and UV ranges.  

For the propagation of UHECR nuclei we used the code described in more detail in Allard et al. (2005,2006), where interactions of protons and nuclei were modeled using a Monte Carlo technique. All the relevant interaction and energy-loss channels were taken into account. Adiabatic losses and pair production were treated as continuous processes. Regarding the pair production we followed the cross section and inelasticities given in Rachen (1996), which we also employ for the scaling of these quantities with the mass and charge of the nucleus. The spectra of the produced pairs on the different photon backgrounds, which are in most cases the strongest contribution to electromagnetic cascades (see below) were calculated following the work of Kelner and Aharonian (2005). The nucleon pion production, which leads to the production of photons, neutrinos and electrons or positrons, was simulated using the SOPHIA event generator (Mucke et al., 2001).  In the case of  nuclei, the different channels of the giant dipole resonance (GDR) were modeled using the theoretical calculations from Khan et al. (2005) and parametrizations from Rachen (1996) for nuclei with mass A$\leq 9$. Cross sections and nucleon yield for the quasi-deuteron and pion production (baryonic resonances) were also treated using Rachen (1996). We also followed his treatment for the nuclear transparency and the kinematics of the produced pions. Our Monte Carlo code allows us to propagate protons or nuclei from their source to the observer, following the evolution of the energy, mass and charge and the production of secondary photons, pairs and neutrinos. Neutrinos are supposed to suffer only from adiabatic losses during their propagation, while $\rm e^+e^-$ pairs and photons eventually initiate electromagnetic cascades. 

\begin{figure}[!t]
\center
\includegraphics[width=1.0\columnwidth]{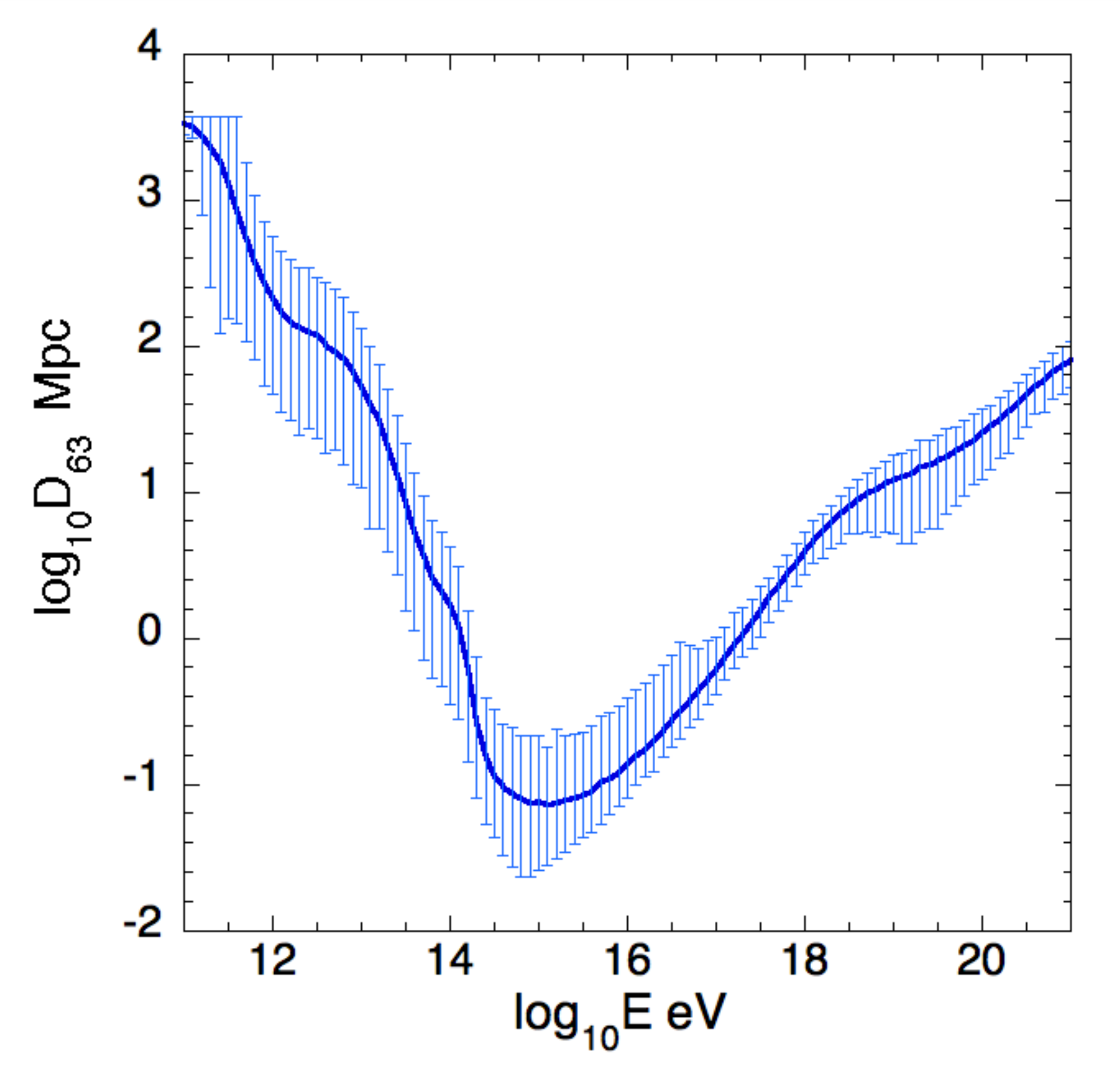}
%\vspace{-1.3cm}
\caption{Effective loss length (i.e., distance needed for the most energetic particle of the electromagnetic cascade to have less than 37\% of the energy of the primary photon) as a function of the energy of the photon that initiated the cascade. The spread (1 RMS) is also displayed. }
%\vspace{-0.6cm}
\label{fig:D63}
\end{figure}

The latter were calculated using a Monte Carlo code computing the most relevant interaction channels (i) for photons: the $\rm e^+e^-$ pair production (Peskin \& Schroeder, 1995) and the double pair production (following Lee, 1998), (ii) for electrons and positrons: the inverse Compton scattering (Jones, 1968; Zdziarski \& Svensson, 1989) and the triplet pair production (Haug, 1975; Mastichiadis, 1991;  Anguelov et al., 1999). The competition between these processes was treated stochastically, whereas the synchrotron and adiabatic losses were treated as continuous processes. Cascades were calculated in a thin grid of electrons and photons with a primary energy between 100 MeV and $10^{21}$ eV  (with a 0.05 logarithmic step) and a grid of redshift between $10^{-5}$ and 8  (with a 0.05 logarithmic step).  {Five-hundred cascades} were calculated for each combination. To limit excessive computation times, we applied a simple thinning algorithm as soon as the cascade particles fell below 0.01\% of the primary energy. When a secondary photon, electron or positron is produced in our UHECR propagation Monte Carlo code, a pre-calculated electromagnetic cascade is randomly chosen at the corresponding energy and redshift and the electromagnetic particles reaching the earth are stored along with cosmic rays and secondary neutrinos. We defined the effective loss length as the mean distance needed for the most energetic particle of the cascade (which is most of the time not the same as the primary particle that initiated the cascade) to fall below 37\% of the energy of the primary particle. The effective loss length was computed for photon-initiated cascades using our Monte Carlo, the result is displayed in Fig.~\ref{fig:D63}. Although numerically different  from the loss length calculated in Lee (1998) because of the different definition used, the same features in the energy evolution are reproduced.  {The successive roles played by the different photon backgrounds in the opacity of the universe to electromagnetic particles are visible}. In the following we assume an extragalactic magnetic field of $10^{-2}$ nG in all our calculations. Although we do not discuss the influence of the field strength in more detail, it was previously shown that fields below 1 nG do not significantly modify the expected photon flux above $10^{19}$ eV (see for instance Gelmini et al., 2008) and we checked that the expected GeV-TeV diffuse flux was not strongly modified by magnetic fields of these magnitudes. We keep in mind, however, that fields stronger than $\sim$1 nG might strongly modify UHE photon fluxes as soon as UHE cascades start developing. 
%Gathering all MFP together, one defines an effective MFP $\lambda_{eff}$ for any particle. For instance, $\lambda_{eff}$ for an electron reads:
%\begin{equation}
%\label{eqn:LambdaEff}
%\frac{1}{\lambda_{eff}}=\sum\limits_{i}^{bg}\frac{1}{\lambda^{i}_{ICS}} + \sum\limits_{i}^{bg}\frac{1}{\lambda^{i}_{TPP}} + \frac{1}{\lambda_{exp}} + \frac{1}{\lambda_{sync}},
%\end{equation}
%where $bg$ refers to the three different backgrounds. 

\section{Principle of the calculation}
In the next sections we calculate secondary photons and neutrino fluxes for different astrophysical models in terms of composition, maximum acceleration energy at the sources, and cosmological evolution of the cosmic ray luminosity.
The cosmic ray spectrum measured by the Pierre Auger Observatory (Abraham et al., 2010) allows us to normalize our UHECR and secondary neutrino and photon fluxes and to choose spectral indices compatible with the overall shape of the UHECR spectrum. In most cases, we will consider maximum energies at the source $\rm E_{max}=Z\times10^{20.5}$ eV (i.e., we assume that protons and nuclei are produced above the pion production threshold) and perform an exponential cut-off of the injection {\it above} this energy. Sources are continuously distributed between a redshift $z=8$ and a minimum distance of $\rm D_{min}=4$ Mpc. We deliberately did not perform a complete statistical analysis to choose these spectral indices, as was done in Ahlers et al. (2010). Indeed, the compatibility of the different models with respect to experimental spectra can slightly differ when one compares them to the Auger or HiRes data. For instance in the framework of the pair production dip model (Berezinsky et al., 2006) good fits to experimental data can be found more easily with the HiRes estimate of the UHECR spectrum than with the Auger estimate. In this context, we simply use the spectral indices that yield the best compatibility with either the Auger or HiRes spectrum. A more detailed study of the compatibility of the different models to experimental data, where a detailed description of the systematic uncertainties of both experiments would be necessary, would not qualitatively change any of the conclusions in our study however.

For our hypotheses on the cosmological evolution of the source luminosity we follow Kotera et al. (2010) and consider, besides a stationary case (hereafter labeled as "uniform"), the cosmological evolution of the source luminosity corresponding to the star-formation rate (SFR) estimated in Hopkins and Beacom (2006) as well as the evolution of powerful radio galaxies (labeled below FR-II) derived from Wall et al. (2005). Other hypotheses on the luminosity evolution were used in Kotera et al. (2010) such as alternative models of the star-formation rate or different estimates of the luminosity evolution of gamma-ray bursts; these cases were yet found to give secondary fluxes very similar to those produced in the SFR case and therefore were not considered in the present study. 

\begin{figure}[!ht]
\center
\includegraphics[width=1.0\columnwidth]{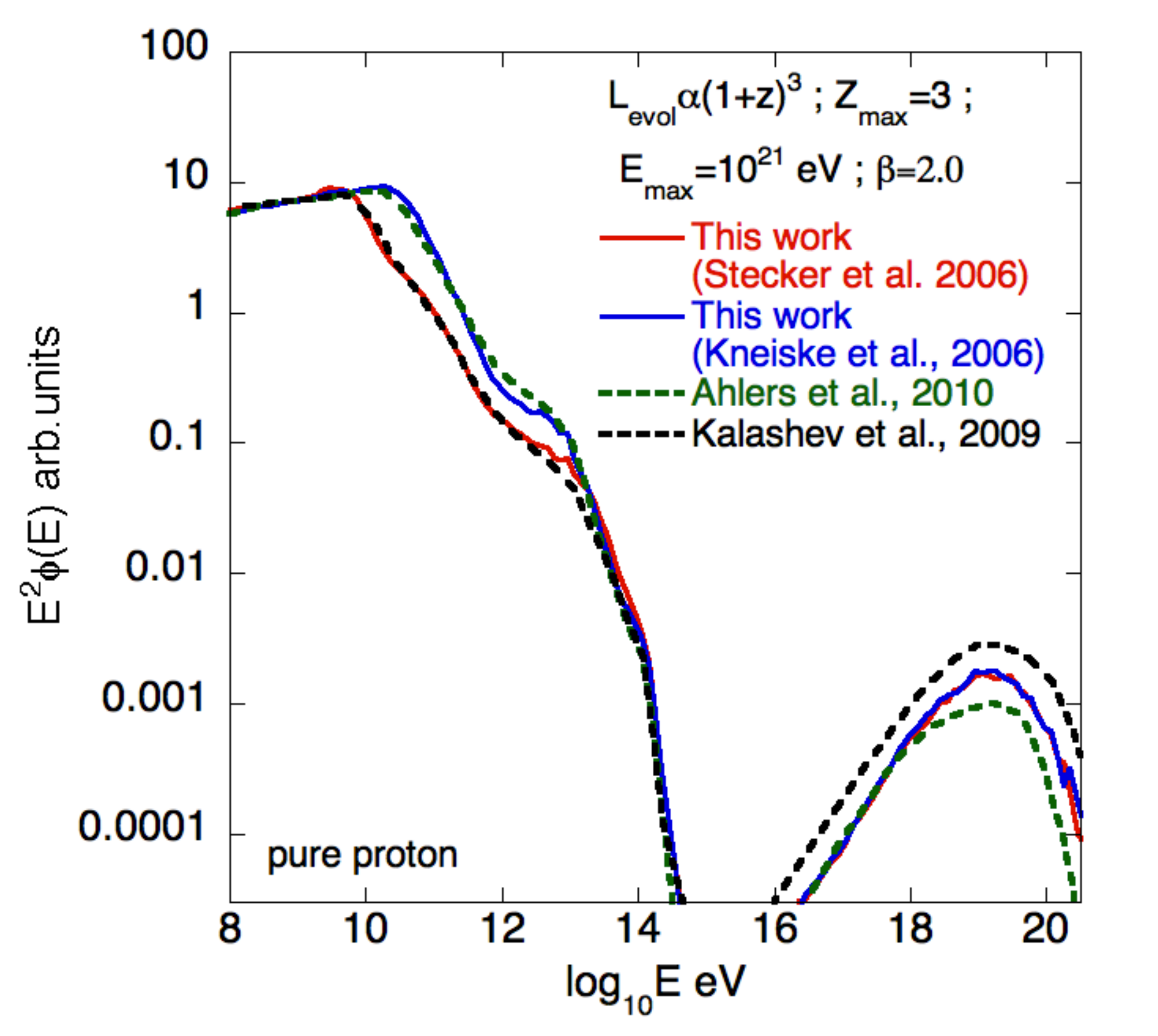}
%\vspace{-1.3cm}
\caption{Expected fluxes of secondary photons for a pure proton composition, a spectral index $\beta=2.0$, a maximum energy $\rm E_{max}=10^{21}$ eV, compared to estimates by Kalashev et al. (2009) and Ahlers et al. (2010). The units are arbitrary but the different fluxes were consistently normalized with respect to their corresponding cosmic ray flux at $10^{19}$ eV (these UHECR fluxes are omitted in this figure, but agree very well). The IR/Opt/UV backgrounds estimated by Kneiske et al. (2004) and Stecker et al. (2006) are used for these comparisons (see text).}
%\vspace{-0.6cm}
\label{fig:comp}
\end{figure}

We compared our numerical model of electromagnetic cascades with previous works by Ahlers et al. (2010) and Kalashev et al. (2009). The comparison is displayed in fig.~\ref{fig:comp} where the photon fluxes are normalized by choosing a similar normalization for the different curves with respect to the corresponding UHECR flux at $10^{19}$ eV. To provide a consistent comparison with the two studies cited above, we used both Stecker et al. (2006) as well as Kneiske et al. (2004) estimates of IR/Opt/UV background photons. None of these backgrounds were used in Ahlers et al. (2010), although their results appear to be very close to those we obtained using Kneiske et al. (2004). The estimate of Stecker et al. (2006) was used in Kalashev et al. (2009). In both cases the comparison agrees very well with previous calculations. One can see the influence of the modeling of the IR/Opt/UV background on the expected photon flux in the GeV-TeV region. The fainter optical and UV background of Kneiske et al. (2004) triggers a pile-up of secondary photons at higher energy than the background of Stecker et al. (2006). The modeling of the optical and UV backgrounds appears to have some influence on the shape of the resulting diffuse gamma-ray fluxes in the GeV-TeV region. Although it is important to stress that these photon backgrounds are not perfectly constrained at the moment, the estimate of Kneiske et al. (2004) shows a much better compatibility with  {the recent Fermi} observations (Abdo et al., 2010). At the highest energies, the modeling of the IR/Opt/UV backgrounds becomes totally irrelevant for secondary fluxes, our simulations agree well with previous works, especially with Ahlers et al. (2010).
 
 \begin{figure}[!ht]
 \center
\includegraphics[width=1.0\columnwidth]{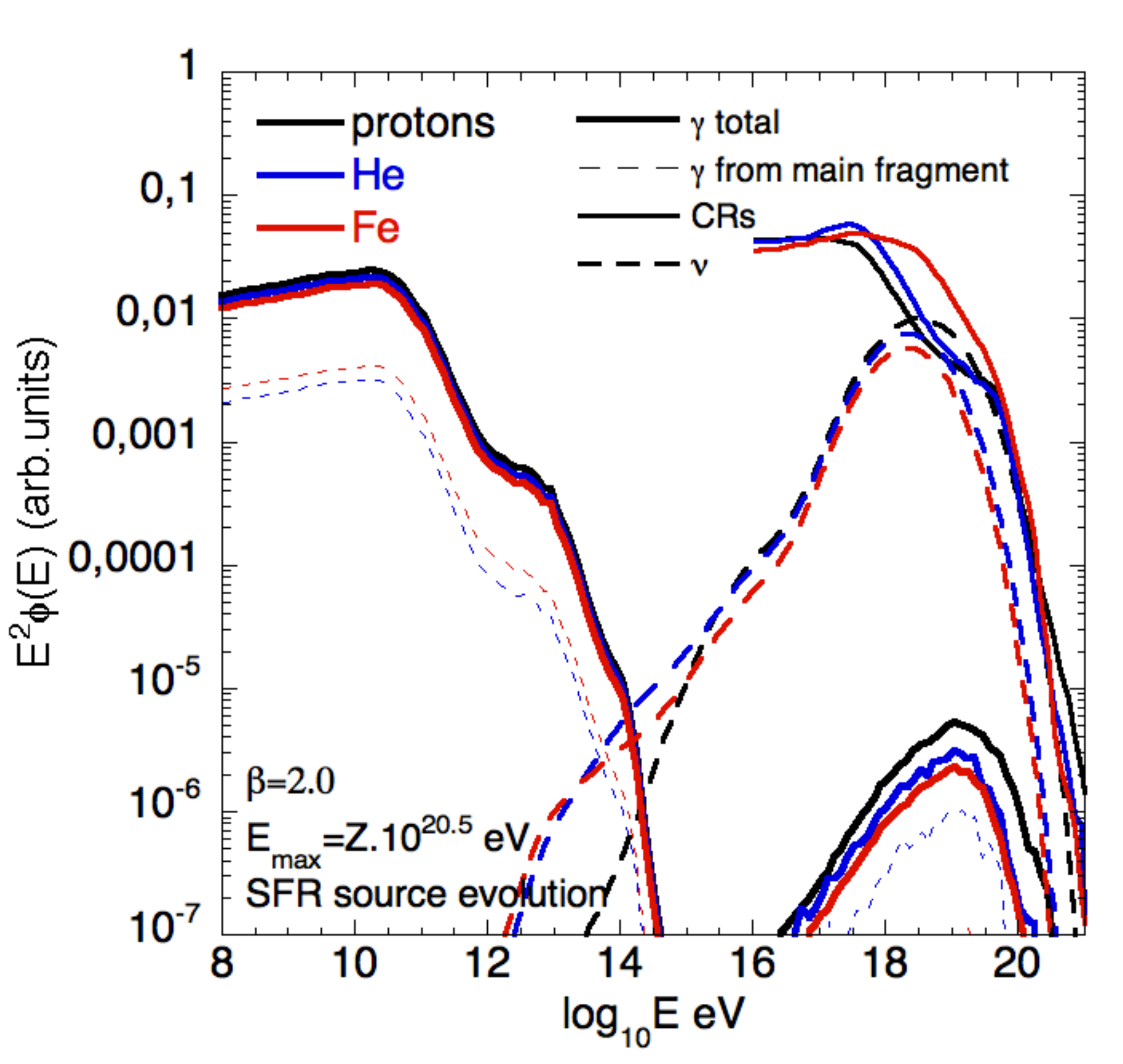}
\hfill
\includegraphics[width=1.0\columnwidth]{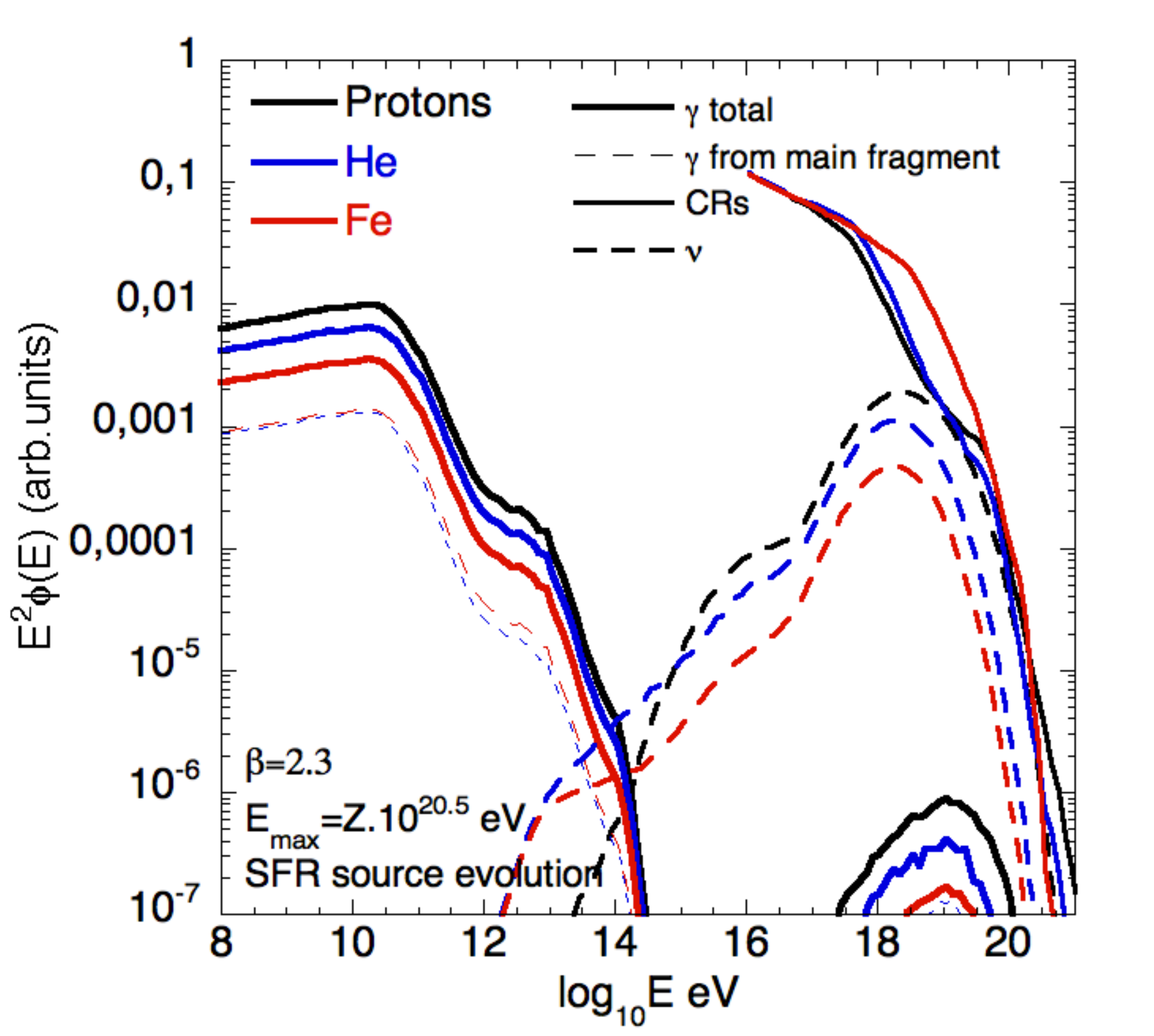}
%\vspace{-1.3cm}
\caption{Top: cosmic ray, neutrino {(summed over all flavors)} and photon spectra assuming three compositions: pure proton, pure helium and pure iron at the source, a source spectral index $\beta$=2.0 and a maximum energy at the source $\rm E_{max}(Z)=Z\times10^{20.5}$ eV. The same cosmic ray luminosity between $10^{16}$ eV and $\rm E_{max}(Z)$ is assumed. The contribution of the main fragment is shown in thin dashed lines. Bottom: same as top panel but assuming a source spectral index $\beta$=2.3.}
%\vspace{-0.6cm}
\label{fig:compnuclei}
\end{figure}

Before presenting the results of our study, we will compare the production of secondaries for different species of UHECR primaries, namely protons, helium, and iron nuclei (we will mainly concentrate on the production of photons because a detailed discussion about neutrinos can be found in Allard et al., 2006, and Kotera et al., 2010).  A comparison between different primary composition is displayed in Fig.~\ref{fig:compnuclei}. One can see the cosmic ray, neutrino and photon outputs for a source luminosity distribution following the SFR for three compositions: pure protons, pure helium, and pure iron.  {The neutrino fluxes shown in all figures are summed over all flavors.} We assume that the same luminosity is injected between $10^{16}$ eV and $\rm E_{max}(Z)=Z\times10^{20.5}$ eV for the three composition models. In the top panel a spectral index $\beta=2.0$ is assumed. The resulting photon spectra appear very similar whatever the primary composition assumed. Indeed, in the case of helium or iron primaries, most of the photon flux is caused by secondary nucleons. Above $\sim2-3\times10^{18}$ eV , secondary nucleons are emitted within a few Mpc from the sources owing to photodisintegration of primary nuclei by CMB photons.  Below the  photodisintegration threshold with CMB photons, nuclei interact with far-infrared photons, although the photodisintegration is much slower (see the evolution of the mean free path in Allard et al., 2006), it remains very efficient on cosmological scales and most of the nuclei emitted by very distant sources release most of their nucleons during propagation. Assuming a complete photodisintegration, secondary nucleons emitted by parent nuclei in the energy range [$\rm A\times E, A\times (E+dE)$] are produced in the energy range [E, E+dE]. Their relative abundance with respect to primary nuclei in the same energy range is $\rm A^{2-\beta}$ and so is the relative total energy they carry. With a spectral index $\beta=2.0$  secondary nucleons   are then as numerous as primary injected nuclei at  the same energy, provided that the parent nucleus is totally photodisintegrated (which is almost always true for nuclei interacting with CMB photons unless the source distance from the observer is below $\sim$ 5 Mpc). Thus, they carry the same total amount of energy as primary protons would do, as soon as the same luminosity is assumed. At the highest energies, the gap between the proton and nuclei cases is mostly due to the fact that we assumed a maximum energy proportional to the charge of the nucleus (to be consistent with the assumption that we make below) and would mostly disappear if we had assumed a scaling with A (note that in this case, the iron case would remain slightly lower owing to the higher absorption probability for the produced pion). The contribution of the residual nucleus to the low-energy photon flux (shown in thin dashed lines) is quite low. This contribution is mainly caused by the pair production mechanism. For nuclei the loss length decreases as $\rm Z^2/A$ at a given Lorentz factor. This process is thus quite efficient for iron nuclei above $\sim6\times10^{19}$ eV (at z=0, see Allard et al., 2006). However, unlike in the proton case, this process has to compete with photodisintegration and is rapidly overwhelmed by GDR interactions with CMB photons (above $\sim3\times10^{20}$ eV for iron nuclei at z=0). As a result, it is only efficient on a short energy range.

As soon as the spectral index becomes softer than 2.0 (see bottom panel of Fig.~\ref{fig:compnuclei} where a spectral index $\beta=2.3$ is considered), the injected luminosity decreases with energy, the pure proton and compound nuclei cases are no longer equivalent and the energy transferred to neutrinos and electromagnetic particles becomes lower as the mass of the parent nucleus increases (according to the scaling in $\rm A^{2-\beta}$) . Because the threshold for the pair production of nuclei is lower than for GDR interactions with CMB photons, the relative contribution of this process (and thus the contribution of the parent residual nucleus) becomes higher for softer spectral indices. Unless assuming very soft spectral indices, expected fluxes of secondaries remain in the same order as in the proton case. A similar discussion can be found in Ahlers and Salvado (2011).
 
As can be seen in Fig.~\ref{fig:compnuclei}, for the same assumed luminosity, the flux of cosmic ray expected on earth is higher in the case of a pure iron composition. This is because protons and then helium have a lower threshold for pair production interactions and as a result suffer faster energy losses than heavier nuclei (at least between $\sim 10^{18}$ and $3\,10^{20}$ eV see Allard et al., 2006 and 2008). Their flux is then more attenuated in this energy range. To reach a certain cosmic ray flux normalization for instance at $10^{19}$ eV, the required luminosity is usually lower if a heavy composition is assumed, and then contributes to the usually lower secondary fluxes found when assuming a heavy composition (see below).
  
\section{Diffuse neutrino and photon fluxes for different astrophysical models}

\begin{figure}[!ht]
\center
\includegraphics[width=1\columnwidth]{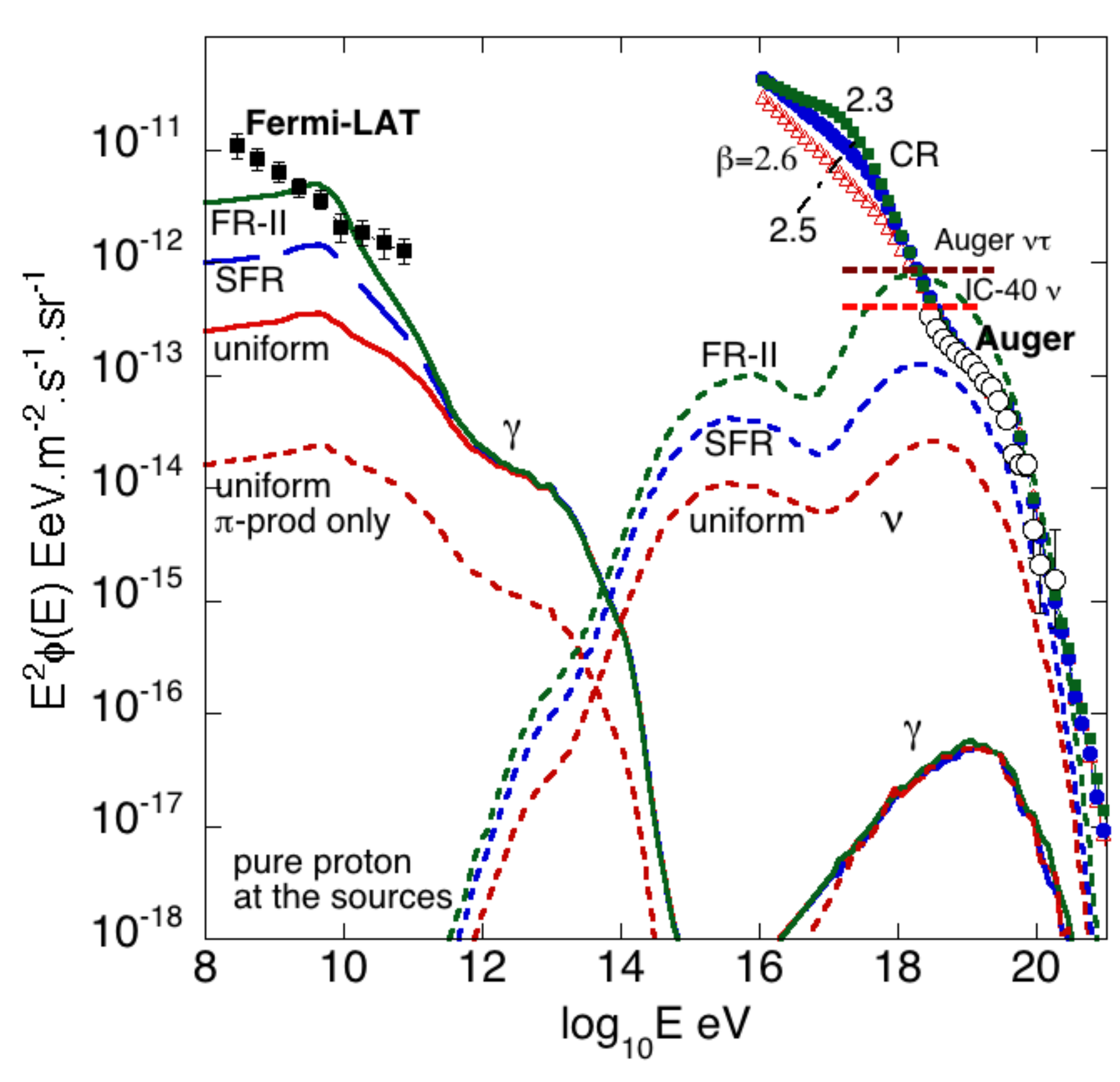}
\hfill
\includegraphics[width=1\columnwidth]{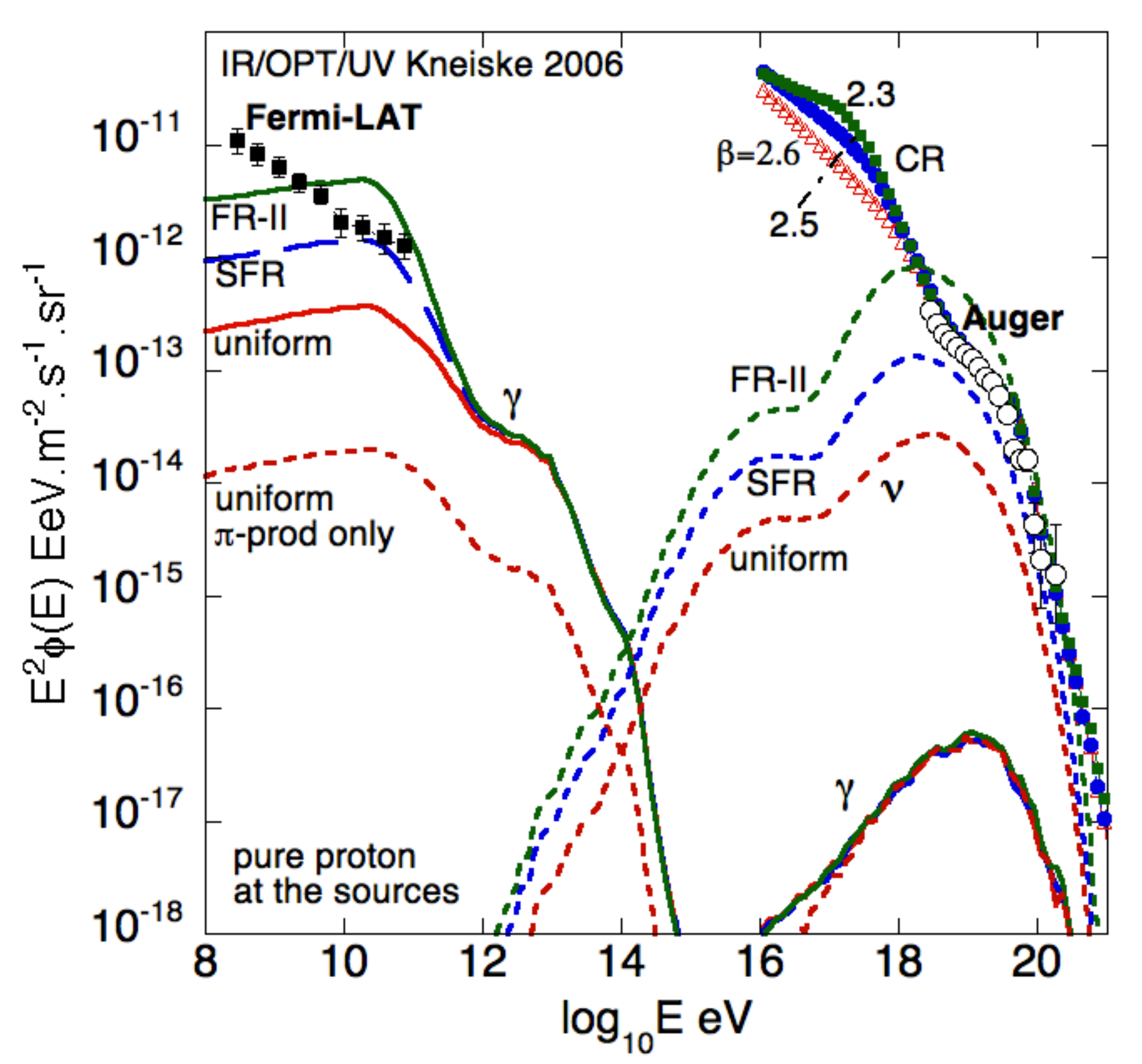}
\caption{Cosmic ray (markers), neutrino (dashed lines) and photon (solid lines) spectra ($\rm E^2\times dN/dE$) for the dip model compared to Auger spectrum (Abraham et al., 2010; open circles) and  {the Fermi }diffuse gamma-ray spectrum (Abdo et al., 2010; black squares). The contribution of the pion mechanism to the photon spectrum is shown (dashed lines). The chosen spectral indices are $\rm \beta=2.6$ for the uniform case (no evolution), 2.5 for SFR and 2.3 for FR-II. The results were computed assuming the IR/Opt/UV background estimate from Stecker et al., 2006 (Top) and Kneiske et al., 2004 (Bottom). In the top panel the Auger 90\% C.L integrated upper limit (2 years) for tau neutrinos assuming a pure $\rm E^{-2}$ neutrino spectrum is also shown for comparison (Abraham et al., 2011; the line represents the central value and was multiplied by 3 assuming a complete mixing of the neutrino flavors). The equivalent IceCube limit (IC-40, red thick-dashed line) is also shown (Abbasi et al., 2011).}
\label{fig:dip}
\end{figure}

In the same spirit as Kotera et al. (2010), we studied different models assuming various compositions and involving different interpretations of the transition from galactic to extragalactic cosmic rays. In all models in this section a maximum energy $\rm E_{max}=Z\times10^{20.5}$ eV is assumed, \emph{i.e.} we consider cosmic rays accelerated above the threshold of pion production with CMB photons. We injected cosmic rays from $10^{16}$ eV to $\rm E_{max}(Z)$ with various spectral indices (choosing for each source evolution hypothesis the spectral index that agrees best with experimental data) and assumed an exponential cut-off of the injection above $\rm E_{max}(Z)$.

\subsection{dip model}

We started our study with the pure proton pair production dip model (Berezinsky et al., 2006). This model is very specific because here the ankle is not related to the transition from galactic to extragalactic cosmic rays, but is instead the signature of proton interactions with CMB photons via the pair production mechanism. The resulting cosmic ray, neutrino and photon spectrum that we obtained for the three hypotheses on the cosmological evolution of the source luminosity  are presented in Fig.~\ref{fig:dip}. Calculations are shown for both estimates of the IR/Opt/UV from Stecker et al. (2006) and Kneiske et al. (2004). Evidently, the photon (at low-energy) as well as the neutrino fluxes are very dependent on the assumed evolution of the source luminosity (see for instance Seckel and Stanev, 2005). Indeed, both low-energy photons and neutrinos come from cosmological distances and can benefit from a lower interaction threshold and denser photon backgrounds at high redshifts. In the case of a strong evolution of luminosity with redshift, the sources with high redshift have a relatively greater weight, which immediately turns into higher secondary fluxes. This is not the case for high-energy photons, because these particles always cascade to low energies unless they are produced in the local universe. The luminosity evolution is thus barely relevant for high-energy gamma-rays, but tends to slightly attenuate the production of high-energy photons because a lower weight is given to local sources. On the other hand, the spectral indices required to fit the experimental data are harder for strongly evolving luminosities with redshift, which means that relatively more luminosity is injected at high energies. In the dip model the two effects compensate almost exactly. Therefore the evolution of the source luminosity has basically no visible effect on the high-energy photon flux. 

When comparing the influence of the IR/Opt/UV model used, one can see that it mainly affects neutrino fluxes in the PeV region. This point was already discussed in Kotera et al. (2010): the fainter photon background in the optical and UV range from the estimate of Kneiske et al. (2004) leads to neutrino fluxes a factor of $\sim2$ lower at $10^{16}$ eV and dropping much faster below this energy. For both of the background models the expected low-energy photon fluxes significantly overshoot the diffuse photon flux measured by Fermi in the scenario of a FR-II evolution of sources. Constraints seem to be more stringent using the photon background by Kneiske et al. (2004), favored by  {the Fermi} observations (Abdo et al., 2010) and in this case the photon flux in the SFR evolution case appears to be very close to  {the Fermi} bounds. Here, we confirm previous results by Berezinsky et al. (2010) and Ahlers et al. (2010), claiming that in the framework of the dip model,  {the Fermi} measurements of the diffuse gamma-ray flux { actually} involve strong limitations on the expected cosmogenic neutrino fluxes.  {By themselves, indeed, ruling out basically all models that yields neutrino fluxes higher than the SFR model, they imply neutrino fluxes almost an order of magnitude lower than the upper limit of the Pierre Auger Observatory (see Abraham et al., 2009; Tiffenberg et al., 2009; Abreu et al., 2001 and Fig.~\ref{fig:dip}) and even lower than the current limits from the IceCube collaboration (Abbasi et al., 2011)}. Constraints obtained from the Fermi measurements can be somewhat dulled by invoking a low-energy cut mechanism\footnote{A  change of the spectral index below $\sim 10^{18}$ eV to a harder value owing to a change of the acceleration regime at the source that allows one to limit the luminosity injected at low-energy, see Berezinsky et al. (2006)} that would leave the UHE neutrino flux unchanged while decreasing the pair production contribution (see below) to the diffuse gamma-ray flux. However, this would be at the expense of the cosmogenic neutrino flux between 1-100 PeV (see Allard et al., 2006).

\subsection{Mixed composition transition models}

\begin{figure}[!ht]
\center
\includegraphics[width=1.00\columnwidth]{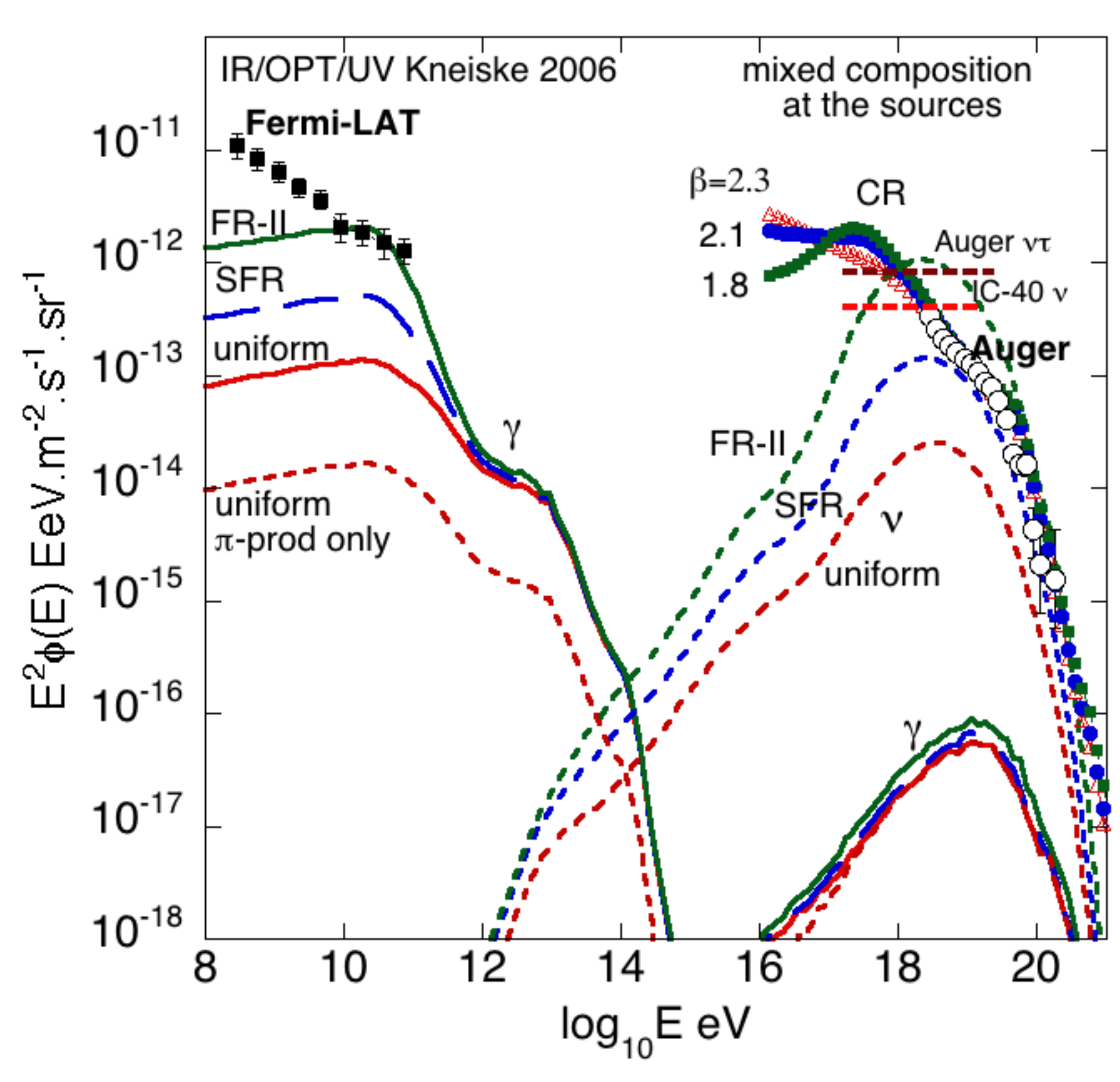}
\hfill
\includegraphics[width=1.00\columnwidth]{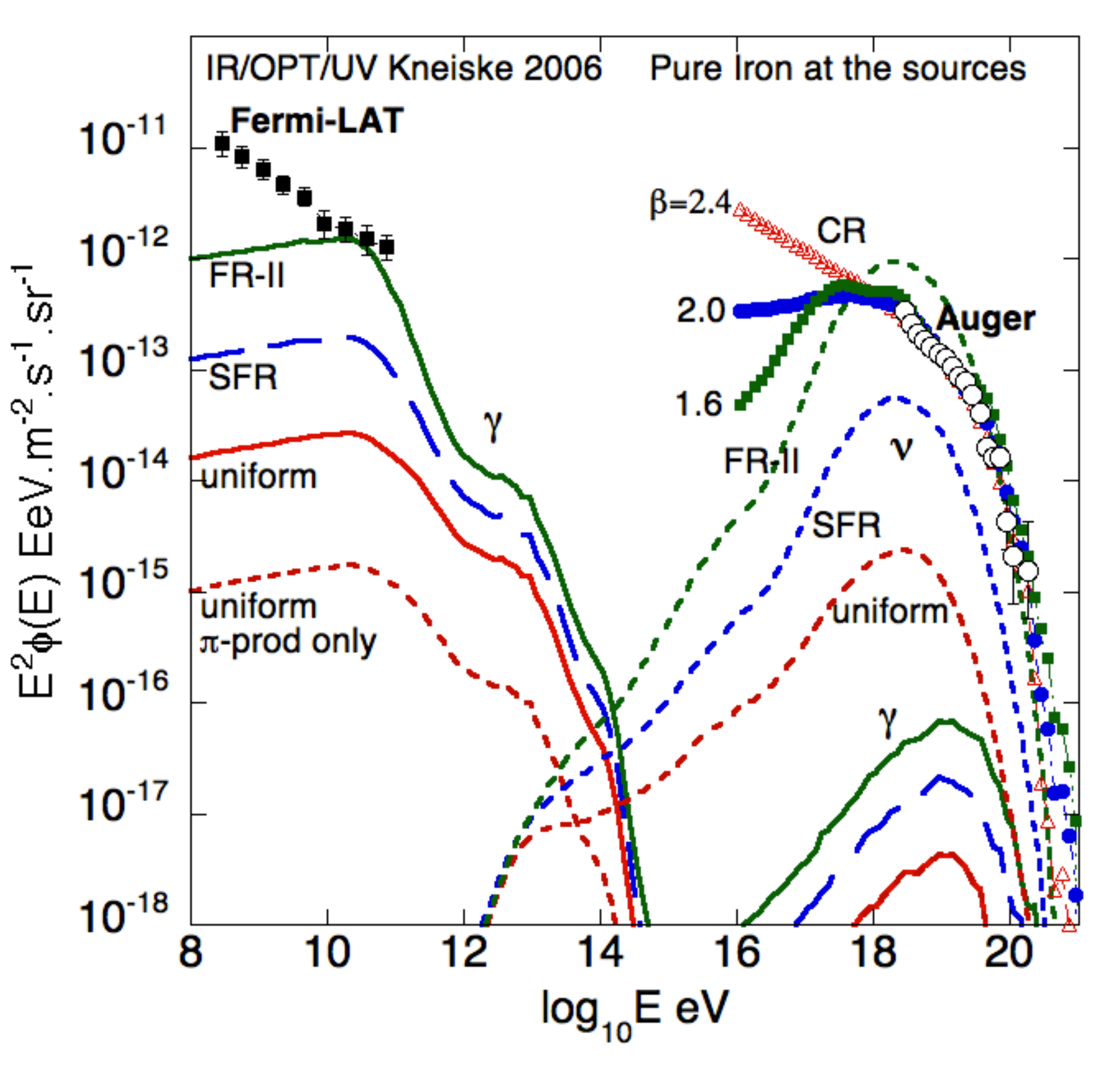}
%\vspace{-1.3cm}
\caption{Same as Fig.~\ref{fig:dip} but considering only the IR/Opt/UV background by Kneiske et al. (2004). Top: We assume a proton-dominated mixed composition and spectral indices $\rm \beta=2.3$ (uniform), 2.1 (SFR), 1.8 (FR-II). Bottom: A pure iron composition is assumed, spectral indices are $\beta=2.4$ (uniform), 2.0 (SFR), 1.6 (FR-II).}
%\vspace{-0.6cm}
\label{fig:mixed}
\end{figure}

We now consider the mixed composition model from Allard et al. (2005). There, the composition at the extragalactic sources is assumed to be similar to that of low-energy galactic cosmic rays. In this case, a pair production dip is no longer possible because of the significant contribution of nuclei to the source composition, and one can fit the cosmic ray spectrum down to the ankle (which is in this case the signature of the end of the transition from galactic to extragalactic cosmic rays) with harder spectral indices than for the dip model. Results are displayed in the top panel of Fig.~\ref{fig:mixed}.  One can see that in this case, as previously shown in Allard et al. (2006) and Kotera et al. (2010), the high-energy neutrino and UHE photon fluxes are very similar to the one obtained for the dip  {model}. At PeV energies, the neutrino fluxes are, however, much lower because of the harder spectral index required to fit the experimental data which leads to lower injected luminosities at low-energy.

The constraints implied by  {the Fermi }diffuse flux appear to be much less stringent for the mixed composition model than for the dip model. Only the FR-II source evolution model seems to be constrained by slightly overshooting  {the Fermi} flux, while the low-energy photons produced in the SFR case are safely below the bounds.  {For the mixed composition model, the bounds given by Fermi are only constraining the most optimistic neutrino flux expectation at the same level as the current upper limits from the Pierre Auger Observatory. The recent upper limit of IceCube clearly rules out the specific FR-II scenario and already constrains less optimistic scenarios between the FR-II and the SFR case. We will return to this discussion below, but we can already point out  that the cosmogenic neutrino fluxes implied by the diffuse gamma-ray flux are strongly model-dependent and  the possibility of observing high cosmogenic neutrino fluxes is not excluded, confirming the conclusion of Ahlers et al. (2010).}

The bottom panel of Fig.~\ref{fig:mixed} shows the case of a pure iron composition at the source. Although very different from the mixed composition case from the point of view of the source composition, the pure iron case has basically the same implications for the interpretation of the ankle. The pure iron case shows stronger variations of the UHE neutrino and photon fluxes with the cosmological evolution assumed than other models. This is caused by a stronger variation of the spectral index required to fit experimental spectra. The uniform case shows a flux of UHE neutrinos and photons much lower than the pure proton case owing to the soft spectral index (2.4) required (although it is harder than 2.6 for the dip model) and, as mentioned in the previous section, the lower luminosity needed for an iron composition (see below). For stronger evolution, harder spectral indices are needed and the gap between the pure proton and pure iron cases becomes smaller (a factor $\sim2$ in the SFR case) and eventually the fluxes are similar for the FR-II evolution where, however, a $\beta=1.6$ spectral index is required. From the point of view of  {the Fermi, Pierre Auger Observatory and IceCube diffuse flux implications,} the pure iron case appears very similar to the mixed composition case.

\subsection{Classic ankle and late-transition models}

\begin{figure}[!ht]
\center
\includegraphics[width=1.00\columnwidth]{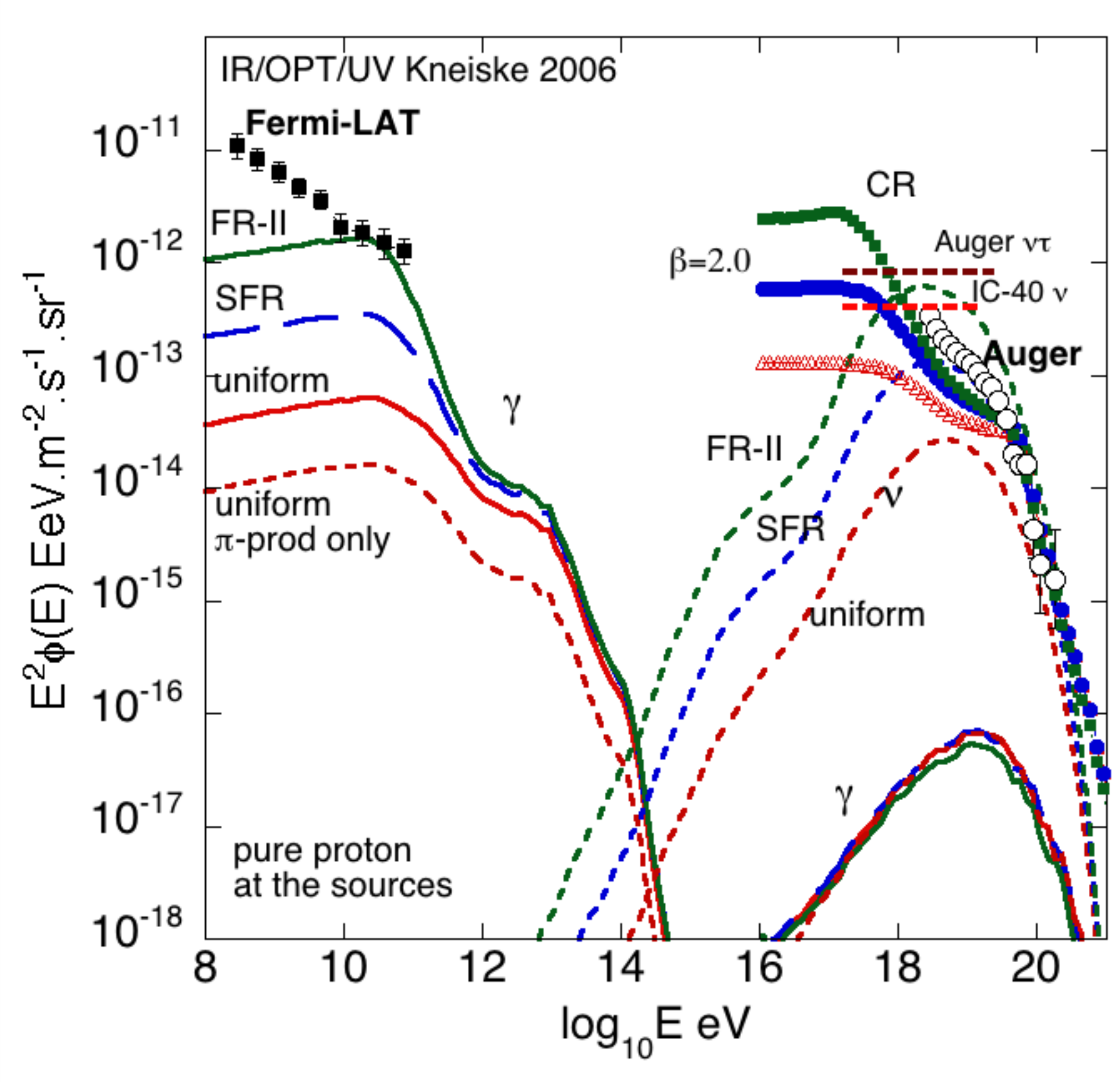}
\hfill
\includegraphics[width=1.00\columnwidth]{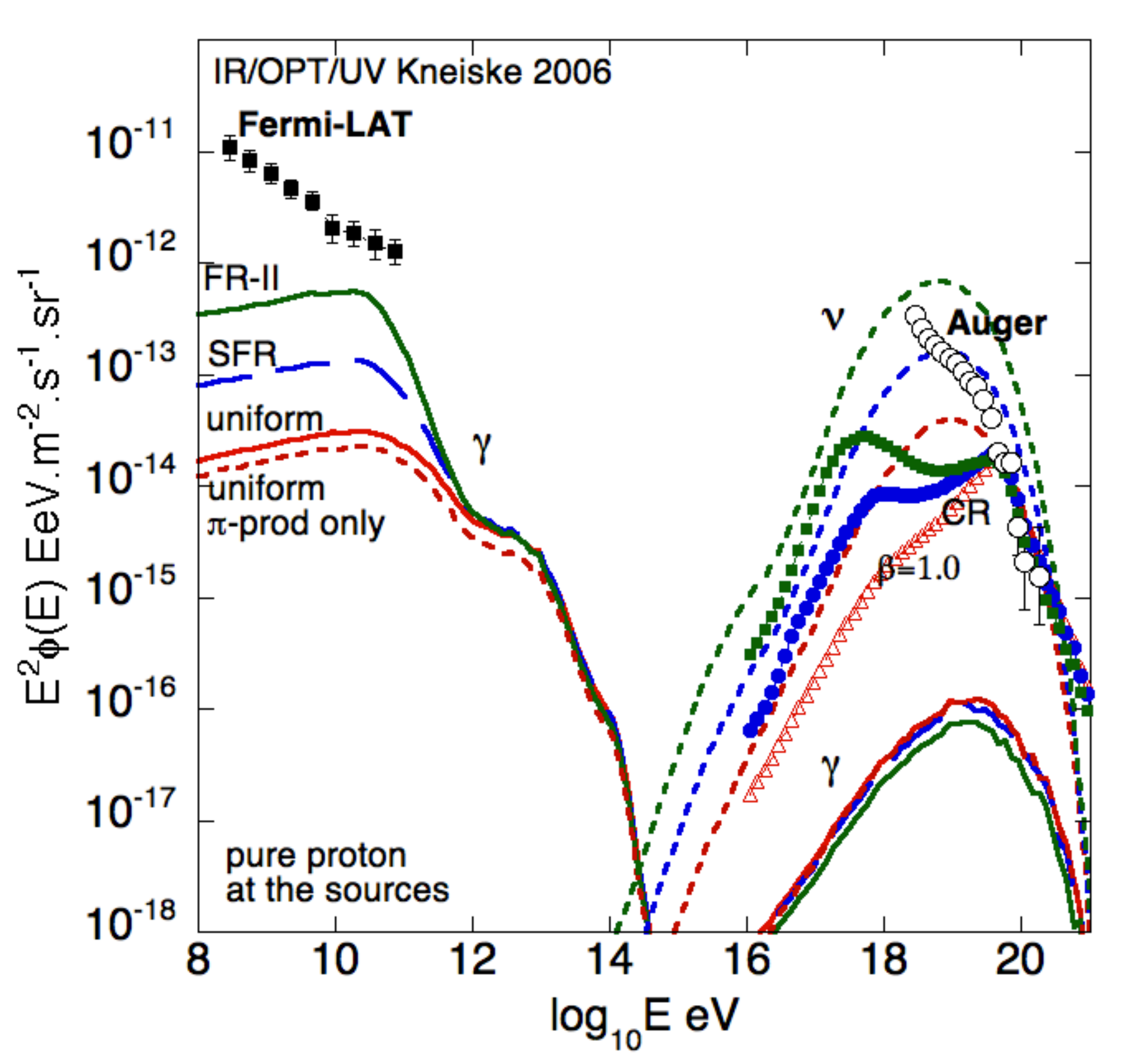}
%\vspace{-1.3cm}
\caption{Same as Fig.~\ref{fig:dip}. Top: a spectral index $\beta=2.0$ is assumed for all source evolution scenarios to reproduce the classic ankle transition model (see for instance Wibig and Wolfendale, 2005). Bottom: a spectral index $\beta=1.0$ is assumed to produce a very late transition best motivated for magnetar-type sources (Epstein et al., 1999; Arons, 2003). }
%\vspace{-0.6cm}
\label{fig:late}
\end{figure}

We also considered later transition models such as the classic ankle model considered for instance by Wibig and Wolfendale (2005). For that purpose a $\beta=2.0$ spectral index is assumed for all evolution models. Results are displayed in the top panel of Fig.~\ref{fig:late}. The experimental spectrum can only be matched above $10^{19}$ eV for this model, and high-energy neutrinos expectations are then not higher than for the dip model (see Takami et al., 2009; Kotera et al., 2010) despite the harder spectral, owing to the lower luminosity needed to match only the highest energy point of the spectrum (see discussion below). The constraints brought by { the Fermi }diffuse flux are then much looser than for the dip model and comparable to those of the mixed composition model.  {A more stringent constraint comes from IceCube which is now able to rule out the highest neutrino fluxes expected in the FR-II scenario.}

Finally, we studied a very late transition model, assuming an extremely hard spectral index $\beta=1.0$ for all evolution scenarios (see the bottom panel of Fig.~\ref{fig:late}). This injection spectrum indices are for instance expected for an acceleration by strong electric fields in the environment of young magnetars (Epstein et al., 1999; Arons, 2003). In this case, this type of extragalactic sources only significantly contributes to the highest energy end of the spectrum and the required luminosity is then very low, and UHE neutrino and photon outputs do not exceed the previous cases. But the  {Fermi diffuse flux does not constrain any of the evolution models by itself, whereas the neutrino fluxes obtained in the FR-II scenario are already constrained by IceCube}.

\section{Observability of diffuse fluxes of UHE neutrinos and photons}

\subsection{Fermi constraints on cosmogenic neutrino fluxes}

\begin{table}[!ht]
\begin{center}
   \caption{\label{table} Cosmic-ray luminosity per unit volume (at z=0, in unit $10^{44}\rm erg\,yr^{-1}\,Mpc^{-3}$) above 10$^{18}$ eV (resp. above $10^{17}$ eV) required for the different models displayed in Figs.~\ref{fig:dip}-\ref{fig:late}. Spectral indices of sources can be found in the corresponding figures.}
\begin{tabular}{|c|c|c|c|}
  \hline
  model & uniform & SFR & FR-II \\
  \hline
  pure proton dip & 11.8 (47.4) &  7.5 (24.0) & 4.24 (9.1) \\
  mixed comp.& 7.1 (15.1) & 4.1 (6.1) & 2.4 (2.8)\\
  pure iron & 4.0 (9.4) & 1.8 (2.2) & 1.6 (1.8) \\
  classic ankle & 2.2 (2.9) & 2.2 (2.9)& 1.6 (2.2)\\
  late transition & 1.1 (1.1) & 0.7 (0.7) & 0.7 (0.7) \\
  mixed low E$_{\textrm max}$ & 5 (7.5) & 2.6 (2.6) & 2.2 (2.2)\\
  \hline
\end{tabular}
\end{center}
\end{table}

\begin{table}[!ht]
\begin{center}
   \caption{\label{tablecas} Cascade energy densities $\rm \omega_{cas}$ (in unit eV$\rm \,cm^{-3}$) for the models displayed in Figs.~\ref{fig:dip} and \ref{fig:mixed}. }
\begin{tabular}{|c|c|c|c|}
  \hline
  model & uniform & SFR & FR-II \\
  \hline
  pure proton dip & $1.3\times10^{-7}$ & $5.0\times10^{-7}$  & $1.7\times10^{-6}$ \\
  mixed comp.& $4.8\times10^{-8}$ & $1.7\times10^{-7}$ & $6.9\times10^{-7}$\\
  pure iron & $9.5\times10^{-9}$ & $6.8\times10^{-8}$ & $5.1\times10^{-7}$ \\
  \hline
\end{tabular}
\end{center}
\end{table}

In the previous section, cosmogenic secondary fluxes for several astrophysical models were calculated. We have shown that as previously pointed out by Ahlers et al. (2010), the implications of {the Fermi} measurements of the diffuse gamma-ray flux are model-dependent. This model dependance can easily be understood by noticing that large parts of the GeV-TeV gamma-ray flux produced by UHECRs are generated by the pair production mechanism. The energy threshold of this process (for the interaction with CMB photons, which is responsible for most of the energy transfer between cosmic rays and secondaries) is much lower than the pion production mechanism that can either produce neutrinos, photons, or electron and positrons. In other words, the neutrino flux is related only to the subdominant pion-induced component of the diffuse gamma-ray flux, which means that similar UHE neutrino fluxes can be associated with a wide range of GeV-TeV gamma-ray flux depending on how high the flux produced by the pair production mechanism is. For a given normalization of the cosmic ray spectrum at $10^{19}$ eV, the contribution of the pair production mechanism to the energy transfer to electromagnetic particles will ultimately depend on the cosmic ray luminosity injected at low-energy, which is related to the spectral index needed for a given model to reproduce experimental data. The very specific interpretation of the ankle, as a consequence of the interaction of protons with CMB photons, requires soft spectral indices for the dip model with respect to other models considered here and thus involves a higher luminosity injected at lower energy (\emph{i.e.} below $10^{19}$ eV). The cosmic ray luminosities above $10^{17}$ and $10^{18}$ eV needed for the different models we considered are summarized in Table.~\ref{table}. Evidently the dip, indeed, does require higher luminosities than other models, especially in the energy range where the pair production mechanism is dominant (e.g above a few $10^{17}$ eV depending on the assumed evolution of the source luminosity). These higher luminosities explain the larger diffuse gamma-ray flux at low-energy and the fact that the dip model is more constrained by {the Fermi} diffuse gamma-ray flux than the other models with a maximum allowed flux almost an order of magnitude lower, unless a low-energy-cut mechanism is assumed. {These previous results can be summarized concisely by comparing the cascade energy density $\rm \omega_{cas}$ (Berezinsky \& Smirnov, 1975) for some of the models we discussed in the previous section (shown in Tab.~\ref{tablecas}) with the maximum cascade energy density $\rm \omega_{cas}^{max}=5.8\times10^{-7}$ $\rm eV\,cm^{-3}$ (Berezinsky et al., 2010) implied by the Fermi measurement of the diffuse gamma-ray flux.} {For models other than the dip model, which yield equivalent UHE neutrino flux expectations for a given evolution scenario, Fermi bounds are only saturated in the FR-II evolution scenario. This means that, there, the constraints obtained from neutrino observations are already more stringent than those deduced from the observations of the diffuse flux of gamma-rays. Hence, the Pierre Auger Observatory and neutrino observatories will certainly keep providing us with interesting constraints in the next few years. In particular, IceCube upper limits will probably reach the level of  the predictions in the SFR source evolution scenario for most of the models we presented in the previous section within $\sim$ 3-4 years.}
{The Fermi constraints can be made more severe however, by identifying significant contributions of astrophysical sources to the diffuse gamma-ray flux. }On this subject, {the Fermi }collaboration (Abdo et al., 2010) recently reported a contribution on the order of $\sim20\%$ dominated by BL Lac objects. Current levels of this contribution do not strongly modify our conclusions yet.

% the FR-II evolution fluxes seem to more or less correspond to the maximum allowed neutrino flux and it could even be larger in the last case we studied (late transition). This means that Auger and neutrino observatories (IceCube, Km3net) should provide stronger constraints in the next few years  than those brought up by the Fermi diffuse gamma-ray flux. 

The discussion on the implications of {the Fermi }diffuse gamma-ray flux is different for cosmogenic neutrinos in the 1-100 PeV range. Indeed, in this energy range cosmogenic neutrinos are mostly produced by interaction of cosmic ray between $10^{17}$ and $10^{19}$ eV with higher energy background (\emph{i.e.}, infrared, optical or ultra-violet). Strong neutrino fluxes also require high luminosities in this energy range (\emph{i.e.} soft spectral indices), then the predicted fluxes for the dip model  are usually much higher than for the other models (see Kotera et al., 2010).  The photon background density evolution with redshift in the IR/Opt/UV range is weaker than for CMB photons, and moreover, spectral indices required to fit the cosmic ray spectrum are becoming harder for scenarios of a strong evolution of the source luminosity. As a result, the difference between the different source evolution scenarios in terms of neutrino flux in the 1-100 PeV range is smaller than for UHE neutrinos. Unlike for UHE neutrinos, the difference between transition models is larger than the difference between source evolution scenario (\emph{e.g.} the neutrino flux for the dip model with SFR evolution is higher than for the mixed composition model and FR-II evolution). Then the highest allowed cosmogenic neutrino fluxes are those predicted for the dip model and the SFR evolution. In this case, the event rate for IceCube  was estimated in Kotera et al. (2010)  between 0.8 and 0.2 events per year using the Stecker et al. (2006) estimate of the IR/Opt/UV background. One would have to decrease this rate by a factor $\sim$2-3 if using Kneiske et al. (2004) results. 

\begin{figure}[!ht]
\center
\includegraphics[width=1.00\columnwidth]{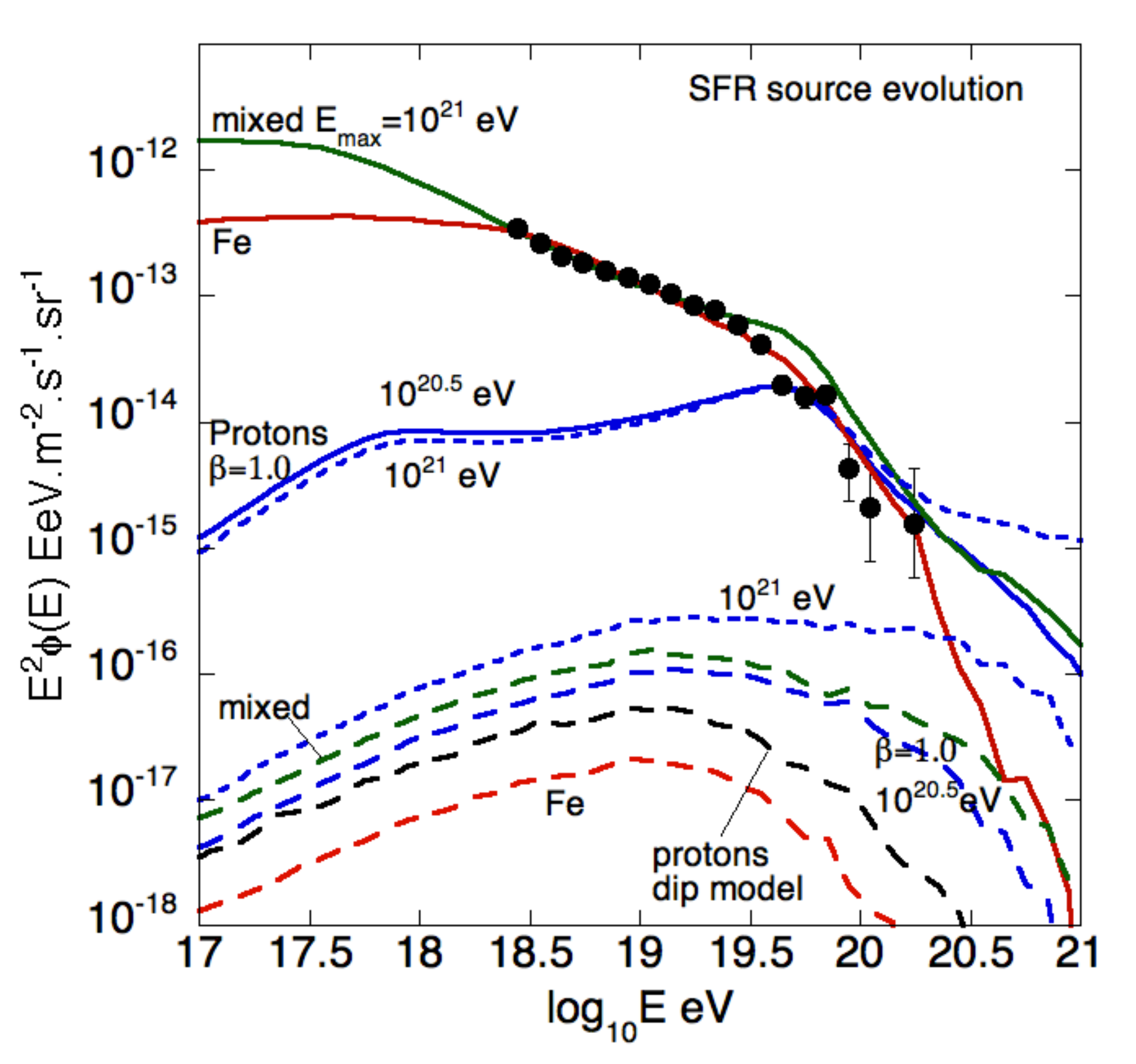}
\hfill
\includegraphics[width=1.00\columnwidth]{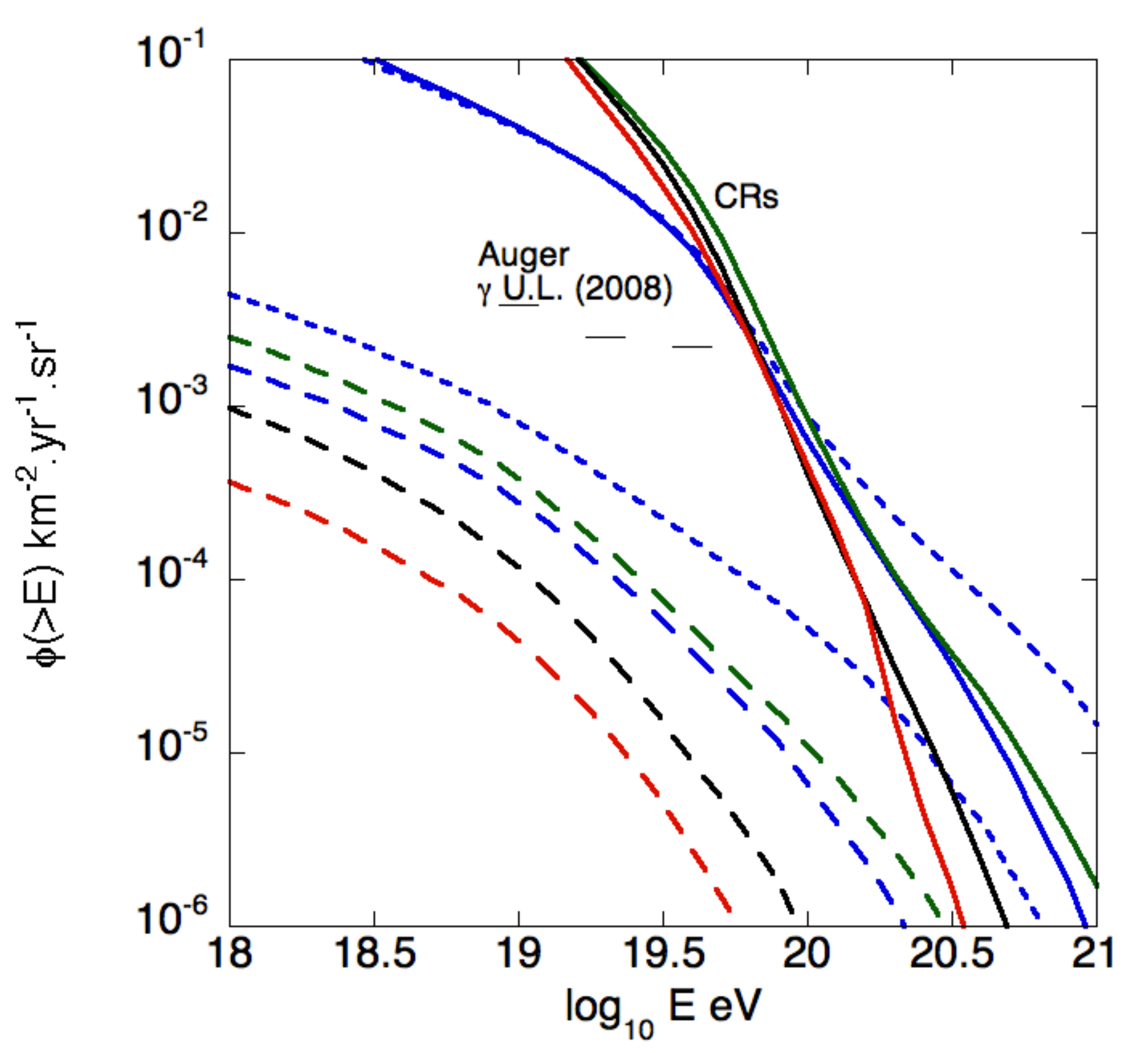}
%\vspace{-1.3cm}
\caption{Top: cosmic ray (solid lines), and photon (dashed lines) spectra ($\rm E^2\times dN/dE$)  compared to Auger spectrum for different models assuming a SFR source evolution : dip model, $\beta=2.5$, $\rm E_{max}=10^{20.5}$ eV; late-transition model, $\beta=1.0$, $\rm E_{max}=10^{20.5}$ eV and $10^{21}$ eV (short dashed lines); mixed composition, $\beta=2.1$, $\rm E_{max}=10^{21}$ eV; pure iron,  $\beta=2.0$, $\rm E_{max}=10^{20.5}$ eV. Bottom: Integrated flux for the same models. Upper limits on the UHE photon flux from the Pierre Auger Observatory are also displayed (Abraham et al., 2008).}
%\vspace{-0.6cm}
\label{fig:summaryphot}
\end{figure}

\subsection{Observability of UHE cosmogenic photons}

Because the cosmological evolution of the source luminosity does not strongly influence UHE cosmogenic photon fluxes (see above), {the Fermi} diffuse flux does not put any constraint on their observability. A non-exhaustive compilation of the UHE photon flux for some of the models presented above is  shown on Fig.~\ref{fig:summaryphot}. The bottom panel shows the integrated fluxes compared to the current upper limits from the Pierre Auger Observatory (Abraham et al., 2008; see also Abraham et al., 2009). The maximum energy was increased to $10^{21}$ eV for the mixed composition (the cosmic ray output then overshoots the data above $3\,10^{19}$ eV) and the late transition models to obtain more optimistic fluxes than in Fig.~\ref{fig:dip}-\ref{fig:late}. The three other models (the second late-transition model with $E_{max}=10^{20.5} eV$, the dip model and the pure iron model) are displayed with {the default} case $\rm E_{max}=10^{20.5}$ eV. Evidently, the predictions displayed here are far below the experimental upper limits. This means that, although these upper limits severely constrained the contribution of most (if not all) Top-Down models, UHE cosmogenic photons would be probably quite difficult to detect anyway (unless in the case of very optimistic models, \emph{e.g.} a late-transition model with $E_{max}>10^{21.5} eV$). These limits are expected to decrease by at least one order of magnitude during the Pierre Auger Observatory operating time (Risse and Homala, 2007). In this context the three most optimistic cases displayed in Fig.~\ref{fig:summaryphot}, corresponding either to very high values of the maximum energy or very hard spectral indices, might be constrained. 

It could be argued that a continuous source distribution is not a realistic modeling for the local universe. For a FR-II type sources, for instance, it is well known that this type of powerful sources is extremely rare in the local universe. The cosmogenic UHE photon flux would then be a lot lower than those displayed in Fig.~\ref{fig:dip}-\ref{fig:late} and the observation of cosmic rays above 3-4$\times10^{19}$ eV would be difficult to explain (owing to the cosmic ray horizon), while the low-energy photon flux and UHE neutrinos would basically remain unchanged. Furthermore, local over-densities (see for instance Gelmini et al., 2007; Taylor and Aharonian, 2010) in the source distribution as well as a nearby source with a high contribution to the total cosmic ray flux (see Taylor et al., 2009 and the next section) can strongly affect the expected UHE photon flux. We study the impact of the contribution of single sources in section~6.  

\subsection{Consequences of the UHE cosmic ray composition}

The discussion on the effect of composition on the expected flux of secondaries is complicated because the gap between the predicted fluxes of UHE neutrinos depends on the spectral index required to fit the spectrum for a given maximum energy per unit charge, which itself depends on the source evolution model assumed. We have shown, however, that UHE photon or neutrino fluxes are quite similar for the pure iron and pure proton cases. In any case, as soon as the maximum energy at the sources is above the pion production threshold (and the source luminosity evolution is non-negligible to provide a hard enough spectral index for a pure iron composition) the discussion of the detectability of UHE secondaries does not severely depend on composition.

\begin{figure}[!t]
\center
\includegraphics[width=1.00\columnwidth]{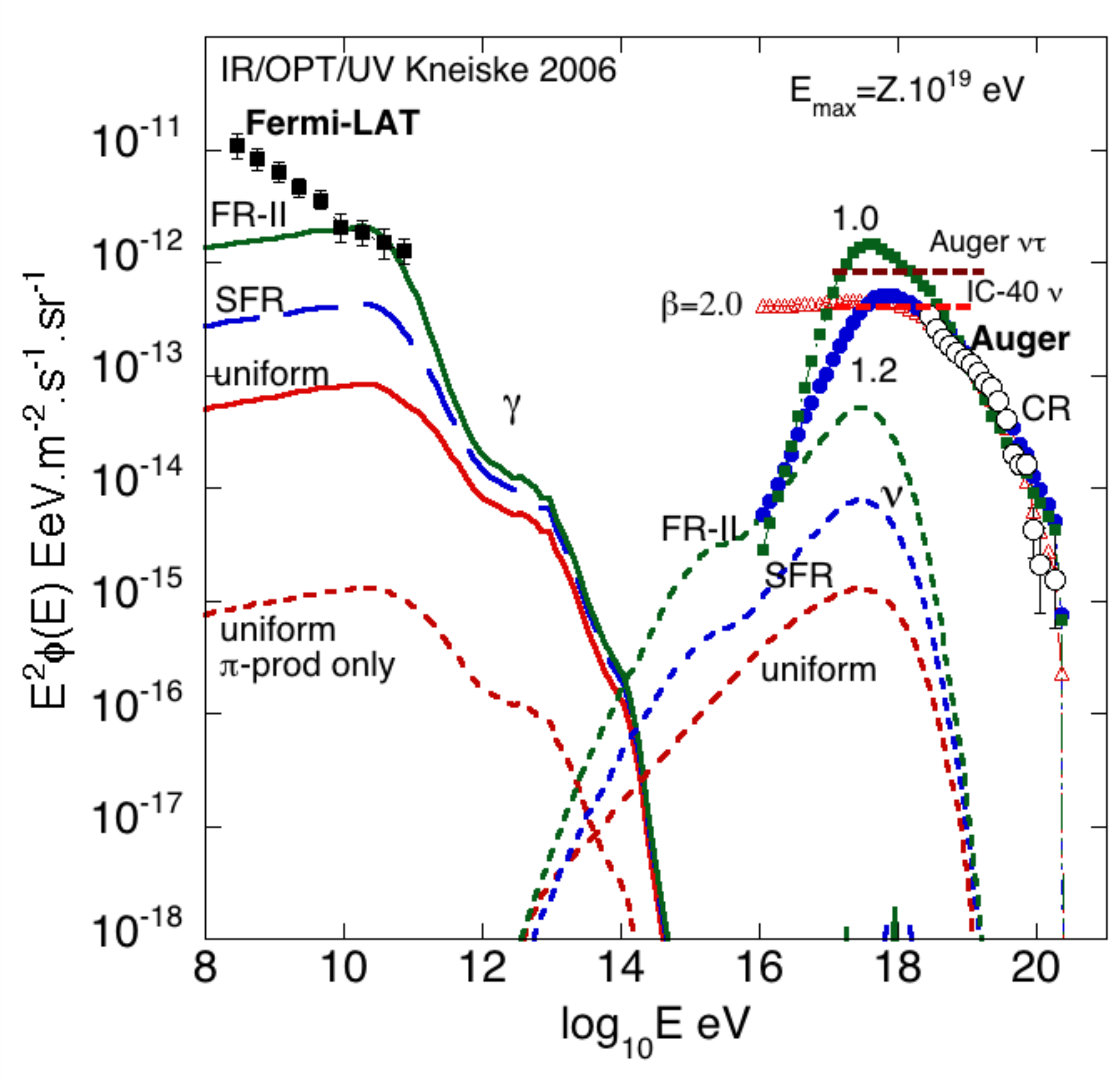}
%\vspace{-1.3cm}
\caption{Same as Fig.~\ref{fig:dip} for the Low $\rm E_{max}$ model (see text).}
%\vspace{-0.6cm}
\label{fig:lowemax}
\end{figure}

The Pierre Auger Observatory recently reported the largest-statistics composition studies above $10^{19}$ eV (Abraham et al., 2010) based on the energy evolution of the maximum of longitudinal development of air shower ($\rm X_{max}$) and its spread.  Obviously, the interpretation of this result in terms of composition of the UHECRs must be considered with care because the hadronic physics taking place at first stages of the shower development is currently not well understood. However, the $\rm X_{max}$ energy evolution behaves as if the composition were gradually becoming heavier with energy, from a light composition around the ankle to a much heavier composition above a few $10^{19}$ eV. As argued in Allard et al. (2008) and Allard (2009), such an evolution of the composition is difficult to justify above $10^{19}$~eV if all the different species present in the source composition are accelerated to the highest energies (above $10^{20}$ eV). In this case, one would expect a composition that becomes lighter (for instance in the case of the galactic mixed composition, see Allard et al., 2007) or more or less steady (low abundances of He and intermediate nuclei compared to heavier nuclei). The most likely explanation of the observed trend involves models where the maximum energy per unit charge is limited around $Z\times10^{19}$ eV or below, \emph{i.e.} models assuming that the proton component is not accelerated at the highest energies. A mixed composition model (enriched in iron to better match the experimental spectrum, Allard et al., 2008; Allard, 2009) with $\rm E_{max}= Z\times10^{19}$ eV (hereafter referred ti as the low-$\rm E_{max}$ model) was then proposed as a possible way to reproduce experimental spectra and the high-energy composition trend. Secondary fluxes for this specific model are presented in Fig.~\ref{fig:lowemax}. Evidently, UHE photon and neutrino fluxes are depleted owing to the lack of cosmic rays accelerated above the pion production threshold and should be well below the detection threshold for all current experiments. 

In this version of the model, however, all sources are assumed to have the same maximum energy, which means none of them is able to accelerate cosmic rays above the pion production threshold. However, one could argue that a dispersion of the maximum energy at the sources (which was proposed by Kachelriess and Semikoz (2006) for instance, to make the dip model compatible with Fermi acceleration motivated spectral indices) is possible (if not likely). This point is usually illustrated by the famous Hillas criterion (Hillas, 1984), stating that the Larmor radius of the accelerated particles cannot exceed the size of the source ($\rm r_L(E)\leq R_s$). One can simply relate the expected maximum energy at the sources with their magnetic luminosity. The simplest estimates (Achterberg, 2002) gives $\rm E_{max}\sim2.5\times10^{20} Z\beta_{s}\Gamma_{s}\times(L_{B}/10^{46}erg\,s^{-1})^{1/2}$ , where $\rm \beta_s$ and $\Gamma_s$ are the speed and the Lorentz factor of the shock (a much more detailed discussion can be found in Lemoine and Waxman, 2009)\footnote{Of course this is a necessary but insufficient condition because it is not dealing with energy losses during the acceleration process. However, adiabatic losses and synchrotron losses would provide larger maximum energy for nuclei than for protons (for synchrotron losses the scaling of $\rm E_{max}$ with A and Z depends on the energy evolution of the acceleration time). For losses caused by to interaction at the sources, this discussion strongly depends on the source environment (see for instance, Allard and Protheroe, 2009)}. From this argument it is often stated that very powerful sources such as FR-II galaxies or GRBs are prime candidates for the acceleration of cosmic ray protons above $10^{20}$ eV. This simple luminosity requirement is much looser for nuclei (especially with large Z), so one could easily conclude that a greater number of sources should be able to accelerate nuclei at the highest energies and that this component might have a dominant contribution to the total UHECR luminosity. {On the contrary}, powerful accelerators might be rare and outnumbered in the local universe or even absent from the UHECR horizon above a few $10^{19}$ eV. Even if the current composition trend shown by the Pierre Auger Observatory is confirmed in the future, one cannot exclude, however, that powerful UHE proton accelerators might be able to contribute at the level of a few percent or a few tens of percent to the cosmic ray flux at $10^{19}$ eV. Figures~\ref{fig:dip}-\ref{fig:late} clearly show that, if these powerful accelerators are strongly evolving, like in the FR-II evolution model, a contribution of 10\% to the total flux at $10^{19}$ eV would be enough to produce a diffuse neutrino flux detectable by IceCube and Auger, which would be a direct signature of the subdominant contribution of these powerful accelerators\footnote{It is often said that the observation of UHE neutrino would invalidate the $\rm X_{max}$ evolution reported by Pierre Auger as being a composition feature and favor an interpretation based on a change of high-energy hadronic physics. Current and future constraints from LHC on high-energy hadronic phenomena taking place in air shower development (d'Enterria et al., 2011) would be crucial to estimate how likely this alternative interpretation is.}. This scenario would produce a diffuse UHE neutrino flux as high as the expectations for the SFR source evolution scenario in the previous section. As a conclusion for this paragraph, even for models with low maximum energies per unit charge (whatever the exact composition, because the absence of UHE secondaries is mainly caused by the low $\rm E_{max}$ and not by the source composition), UHE neutrino observations are still plausible and future more stringent upper limits will be critical to constrain the contribution of powerful accelerators.

\section{Photons and neutrinos as signatures of powerful cosmic ray accelerators}

\begin{figure}[!t]
\center
\includegraphics[width=1.0\columnwidth]{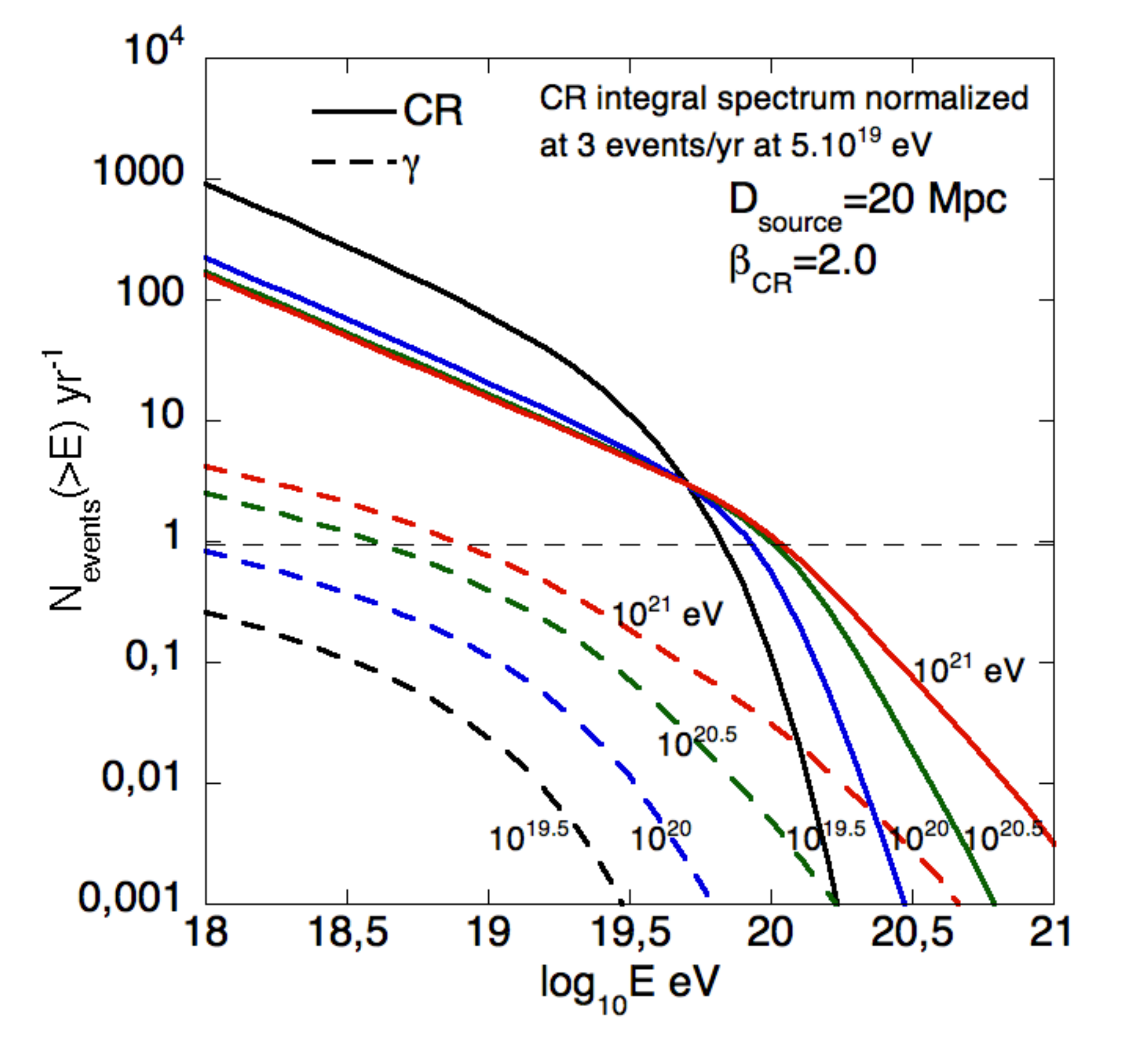}
\hfill
\includegraphics[width=1.0\columnwidth]{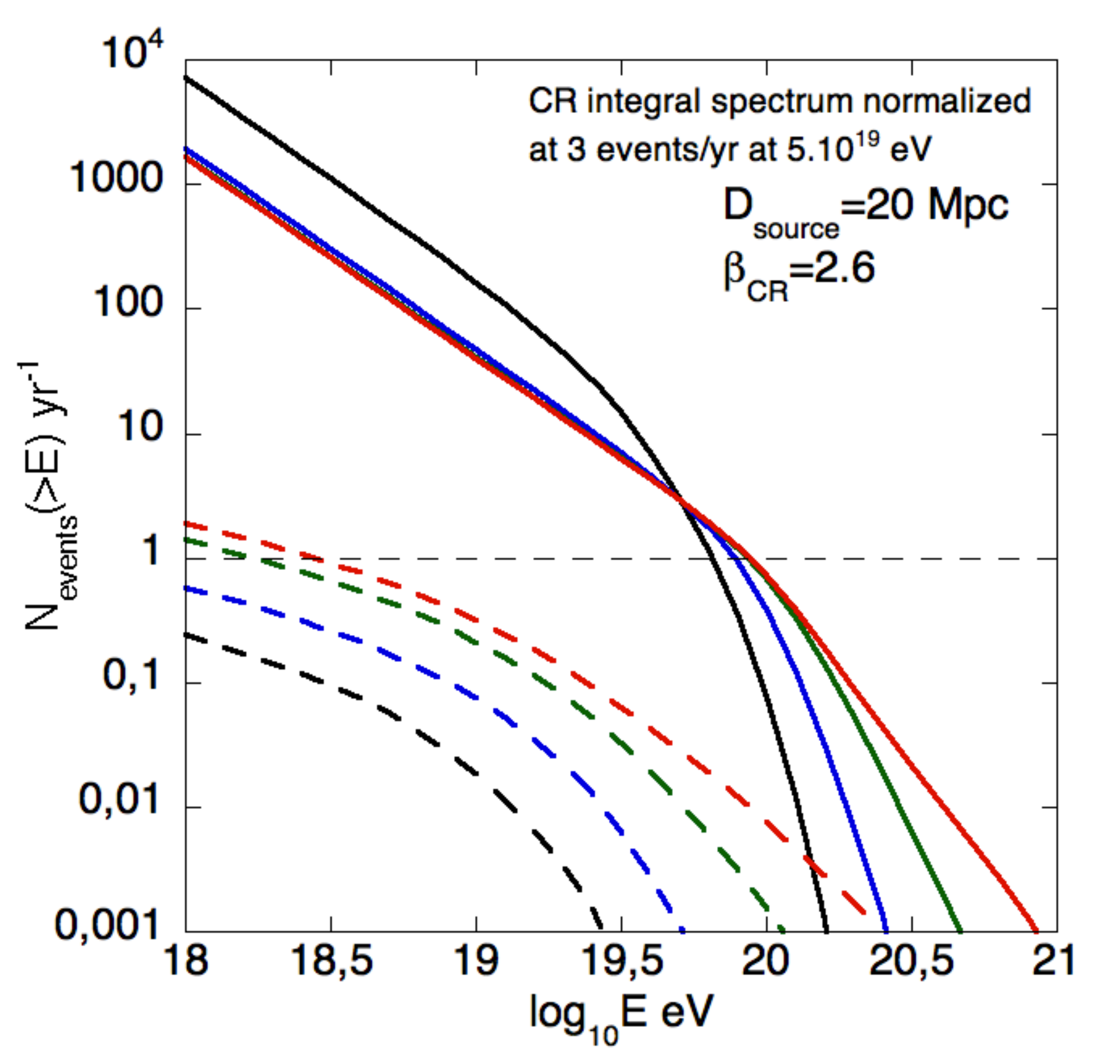}
%\vspace{-1.3cm}
\caption{UHECR and UHE photon-integrated fluxes for a source located at 20 Mpc and different values of the maximum energy: $10^{19.5}$, $10^{20}$, $10^{20.5}$ and $10^{21}$ eV. Top: a source spectral index $\beta=2.0$ is considered. Bottom: $\beta=2.6$, the integrated flux is normalized assuming that the source provides three cosmic ray events above $5\times10^{19}$ eV.}
%\vspace{-0.6cm}
\label{fig:source20}
\end{figure}

Independently of the global composition at the highest energies, the detection of individual sources accelerating protons above $10^{20}$ eV would represent a major step forward for understanding the origin of UHECRs. Cosmic ray observatories are of course prime candidates for that purpose with the measurement of cosmic ray arrival directions at the highest energies. Current statistics is not large enough to firmly establish the detection of significant clustering caused by to the contribution of individual sources, but the next years should allow the Pierre Auger Observatory and JEM-EUSO (Medina-Tanco et al., 2009) to provide more detailed and extended statistics measurements. Above $\sim3\times10^{20}$ eV heavy nuclei are expected to disappear from the cosmic ray composition (provided the closest source of UHECR is not much closer to the Earth than Centaurus A and the abundance of nuclei heavier than iron is low), owing to interactions with CMB photons (Allard et al., 2007; 2008). Cosmic rays at these energies (if any) should be widely dominated by protons or very light fragments (\emph{i.e.}, particles with very high rigidities). This energy range, which should be probed and constrained by JEM-EUSO, is then very promising to observe cosmic rays that are almost guaranteed to point back close to their source. As a consequence of the cut-off of heavy nuclei, however, the cosmic ray flux might be very low if cosmic rays are dominated by heavy nuclei below $3\times10^{20}$ eV. 

\subsection{Cosmogenic UHE photons from single sources}

\begin{figure}[!t]
\center
\includegraphics[width=1.0\columnwidth]{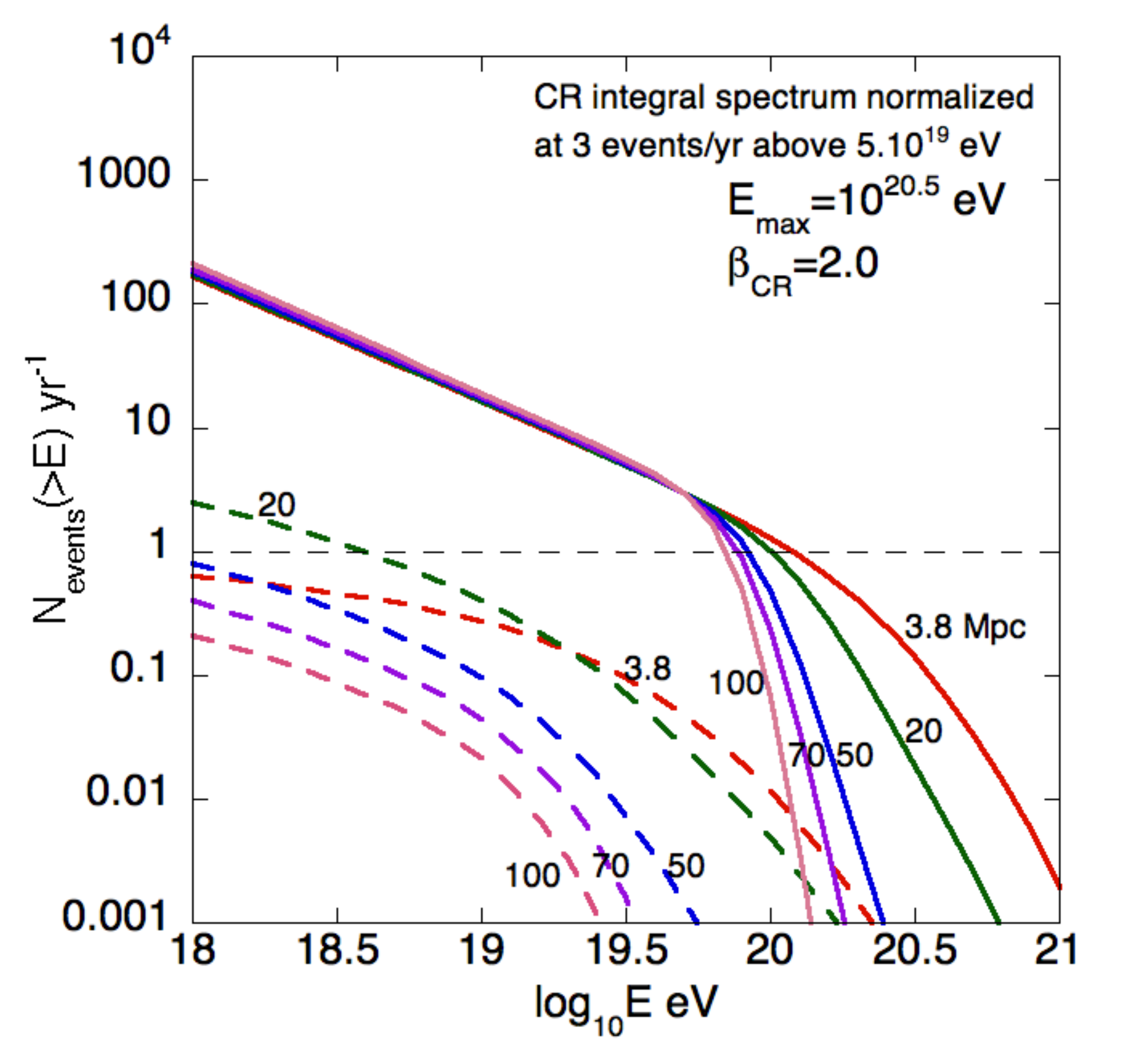}
%\vspace{-1.3cm}
\caption{Same as Fig.~\ref{fig:source20}, but we consider here different source distances (3.8, 20, 50, 70, 100 Mpc), the maximum energy is fixed at $\rm E_{max}=10^{20.5}$ eV and the spectral index $\beta=2.0$.}
%\vspace{-0.6cm}
\label{fig:sourcedist}
\end{figure}

Cosmic ray observatories are also expected to spot individual sources by observing UHE photons. As mentioned above, UHE photon flux is strongly influenced by the distribution of sources in the local universe. Therefore, we studied, as in Taylor et al. (2009), the UHE photon flux of a source accelerating protons at UHE and located at various distances, with various hypotheses on the maximum energy. The example of a source located at 20 Mpc and different values of the maximum energy is shown in Fig.~\ref{fig:source20}, where we assumed that the source was providing three events per year above $5\times10^{19}$ eV (corresponding to a bit less than 10\% of the flux seen by the Pierre Auger Observatory). Obviously, for a spectral index $\beta=2.0$ and maximum energies above $10^{20.5}$ eV the rate of expected events (these are incoming events and not events detected and identified as photons) is between $\sim$0.4 and 1 per year above $10^{19}$ eV and between 0.08 and 0.2 above $10^{19.5}$  eV, which is almost as high as the diffuse flux shown in Figs.~\ref{fig:dip} and \ref{fig:summaryphot} for the dip model. The flux becomes obviously lower for softer spectral indices.
Fig.~\ref{fig:sourcedist} shows the influence of the source distance on the UHE photon flux; the expected flux appears to drop significantly for D$\geq$50 Mpc compared with closer sources owing to the cascading of the electromagnetic particles to lower energies. In contrast, for very nearby sources (for instance, we choose 3.8 Mpc, the distance of Centaurus A) the flux above $10^{19}$ eV is on the same order as in the 20 Mpc case, but harder because of the shorter propagation time that shifts interacting protons to higher energies and prevent the cascades from developing too much toward lower energies.

As commented above, if the source is close enough and has a significant contribution to the total UHECR flux ($\sim10\%$ above $5\times10^{19}$ eV) its contribution to the UHE photon flux can be as high as the whole diffuse photon fluxes calculated in Fig.~\ref{fig:dip}-\ref{fig:late}, above $10^{19}$ eV, provided its maximum energy is around $10^{20.5}$ eV or above (a relatively hard spectral index would also help). For a proton-dominated composition the presence of one or a few of this type of sources could greatly facilitate the detection of UHE photons compared to the default case of a continuous source distribution that we studied. In a low $\rm E_{max}$ type scenario, the detection would be more difficult because a dominant contribution of {sources} with low values of the maximum energy would not provide a background of UHE photon that can be added to that of the single powerful accelerator and a higher contribution of the individual proton source(s) would be needed to compensate for the diffuse flux. A detection of UHE photons would be especially precious because it would prove an UHE accelerator above $\sim10^{20}$ eV per nucleon (which is currently not obvious) and it would be of special interest to compare their arrival directions with those of the highest energy cosmic rays. 

\subsection{Distant point sources of cosmogenic UHE neutrinos and GeV-TeV gamma-rays}

\begin{figure}[!t]
\includegraphics[width=1.0\columnwidth]{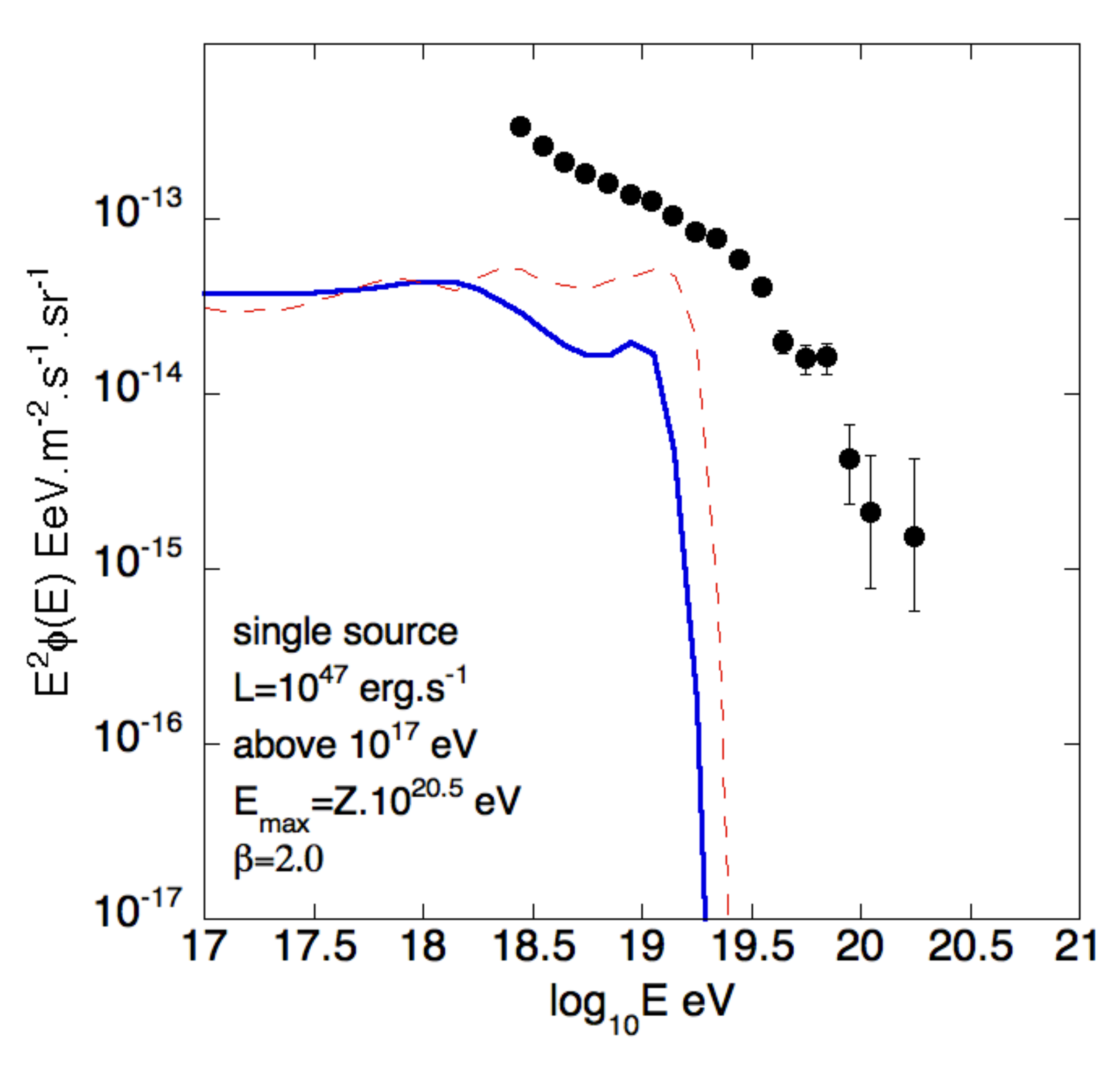}
\hfill
\includegraphics[width=1.0\columnwidth]{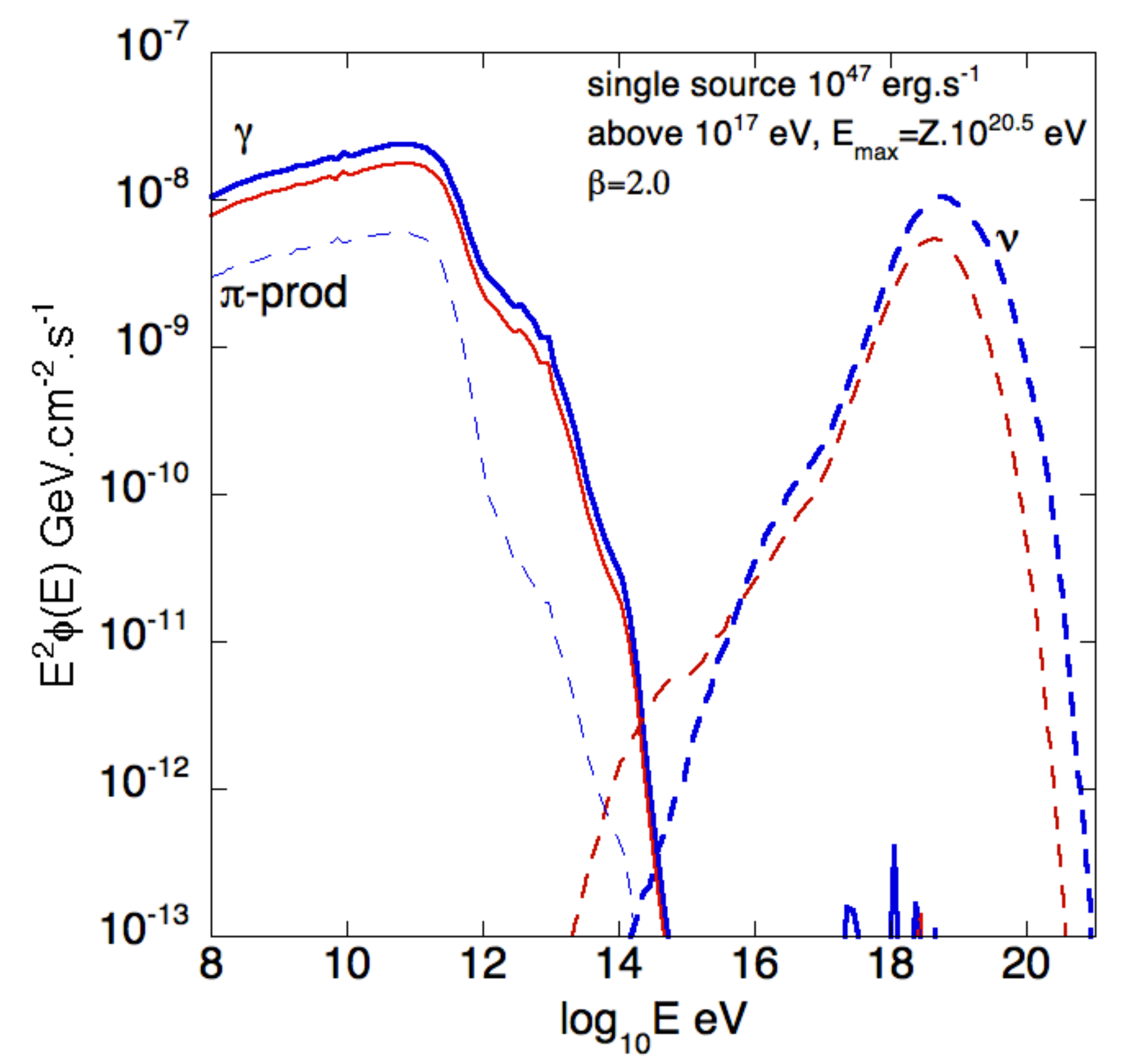}
%\vspace{-1.3cm}
\caption{Top: Cosmic ray spectrum at the Earth for a source located at 1 Gpc emitting a cosmic ray luminosity of $10^{47}$ $\rm erg\,s^{-1}$ between $10^{17}$ eV and $\rm E_{max}=Z\times10^{20.5}$ eV assuming a pure proton (blue solid line) and a pure iron composition (red dashed line) compared with the Pierre Auger Observatory spectrum. Bottom: Cosmogenic neutrinos (dashed lines) and photons (solid lines) for the pure proton (blue) and the pure iron (red) cases corresponding to the same cosmic ray source.}
%\vspace{-0.6cm}
\label{fig:source1000}
\end{figure}

As mentioned above, most powerful accelerators could be located well beyond the cosmic ray horizon (this could be true even if the cosmic ray composition is proton-dominated at the highest energies). The only signature of the cosmic ray acceleration taking place in these sources would be the observation of high-energy neutrinos or sub-TeV gamma-rays that can propagate on cosmological distances, unlike UHE photons or cosmic rays. Studies on the subject have been most recently undertaken by Essey et al. (2010) and Ahlers and Salvado (2010) for nuclei. One potential problem of this type of detection is that distant sources have to be extremely powerful to provide detectable fluxes (see below) at least for cosmogenic neutrino. As an illustration, the example of a source at 1 Gpc with a cosmic ray luminosity of $10^{47}$ $\rm erg\,s^{-1}$ between $10^{17}$ and $\rm E_{max}=Z\times10^{20.5}$ eV  is displayed in Fig.~\ref{fig:source1000} for a pure proton and a pure iron source composition. The top panel shows  the contribution of this source to the total cosmic ray spectrum assuming a completely rectilinear propagation to ensure the very high luminosity assumed does not overshoot the UHECR flux (however, large powers of this magnitude are known for powerful astrophysical objects in other wavelengths). Obviously, the source contributes $\sim15\%$ at $\sim10^{19}$ in the pure proton case, before a cut-off caused by interaction with CMB photons, whereas the iron source contributes at the level of 30\%, their existence is then not constrained by the UHECR cosmic ray spectrum. 

The corresponding cosmogenic neutrinos and photons are displayed in the bottom panel of Fig.~\ref{fig:source1000}. For such a high luminosity the neutrino flux can reach $\rm 10^{-8}\,GeV\,cm^{-2}\,s^{-1}$ at the peak around $3\,10^{18}$ eV. A flux this high might be detected in the future depending on IceCube (Abbasi et al., 2009) or the expected Km3Net sensitivity to point sources at very high energies.  For lower energies the neutrino flux decreases rapidly and the detection of such a signal in the 1-100 PeV range would require a luminosity much higher (see Essey et al., 2010), especially when using the Kneiske et al. (2004) estimate of the IR/Opt/UV backgrounds. The detection of a UHE neutrino signal would provide an important non-ambiguous proof of the existence of very powerful UHE accelerators (\emph{i.e.} acceleration above the pion production) even though the similar flux obtained for a pure iron composition, assuming the same luminosity and same maximum energy per unit charge, shows that these fluxes are not sensitive to the source composition (at least not for hard spectral indices, see discussion above). Moreover, these neutrinos are produced in close vicinity to the source (at this redshift mostly within 10 Mpc) and are then very likely to form a point-like neutrino source for neutrino observatories. This is expected to remain true for beamed sources where the luminosity mentioned above would become the isotropic equivalent luminosity. 

\begin{figure}[!t]
\center
\includegraphics[width=1.0\columnwidth]{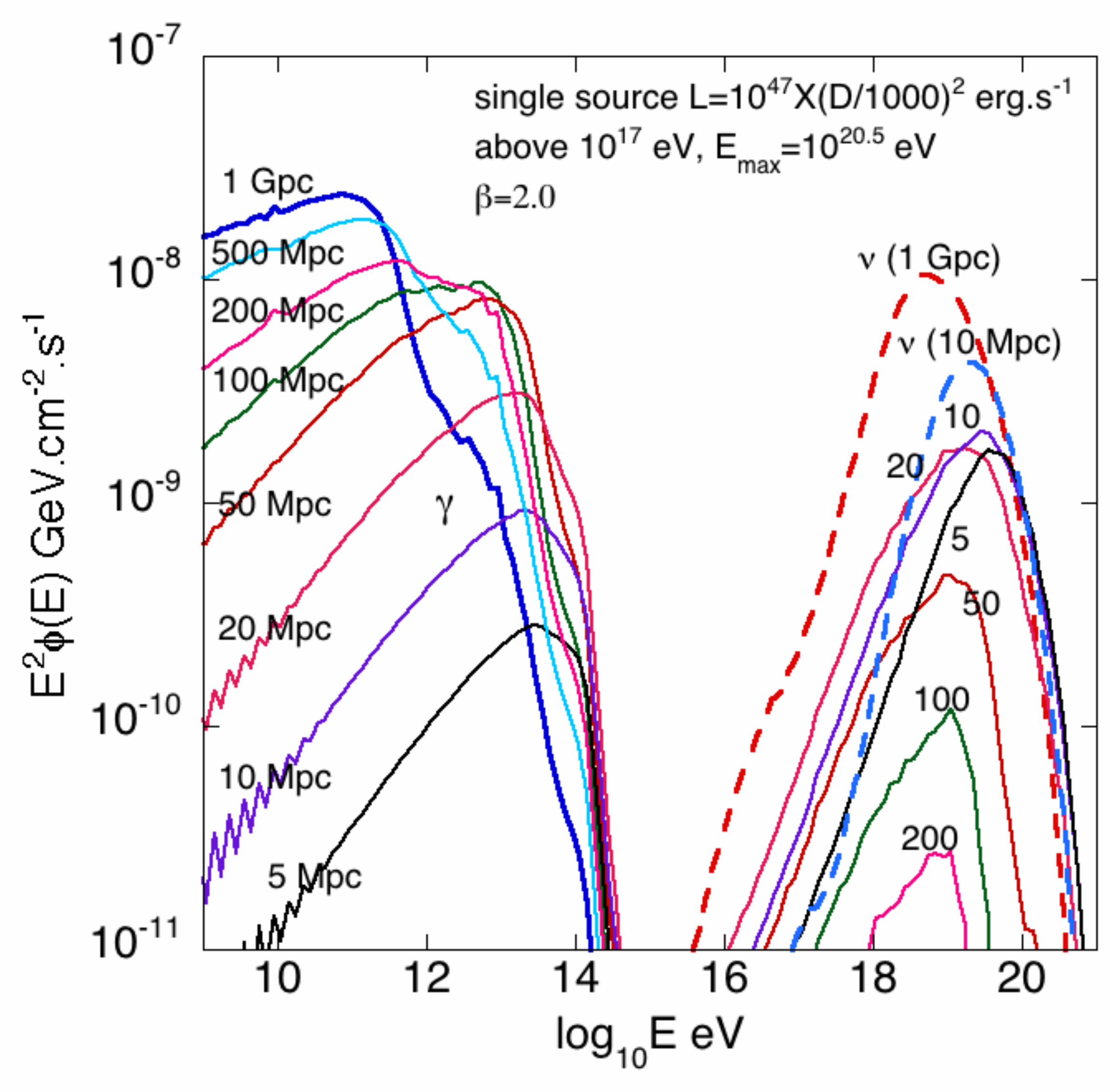}
\hfill
\includegraphics[width=1.0\columnwidth]{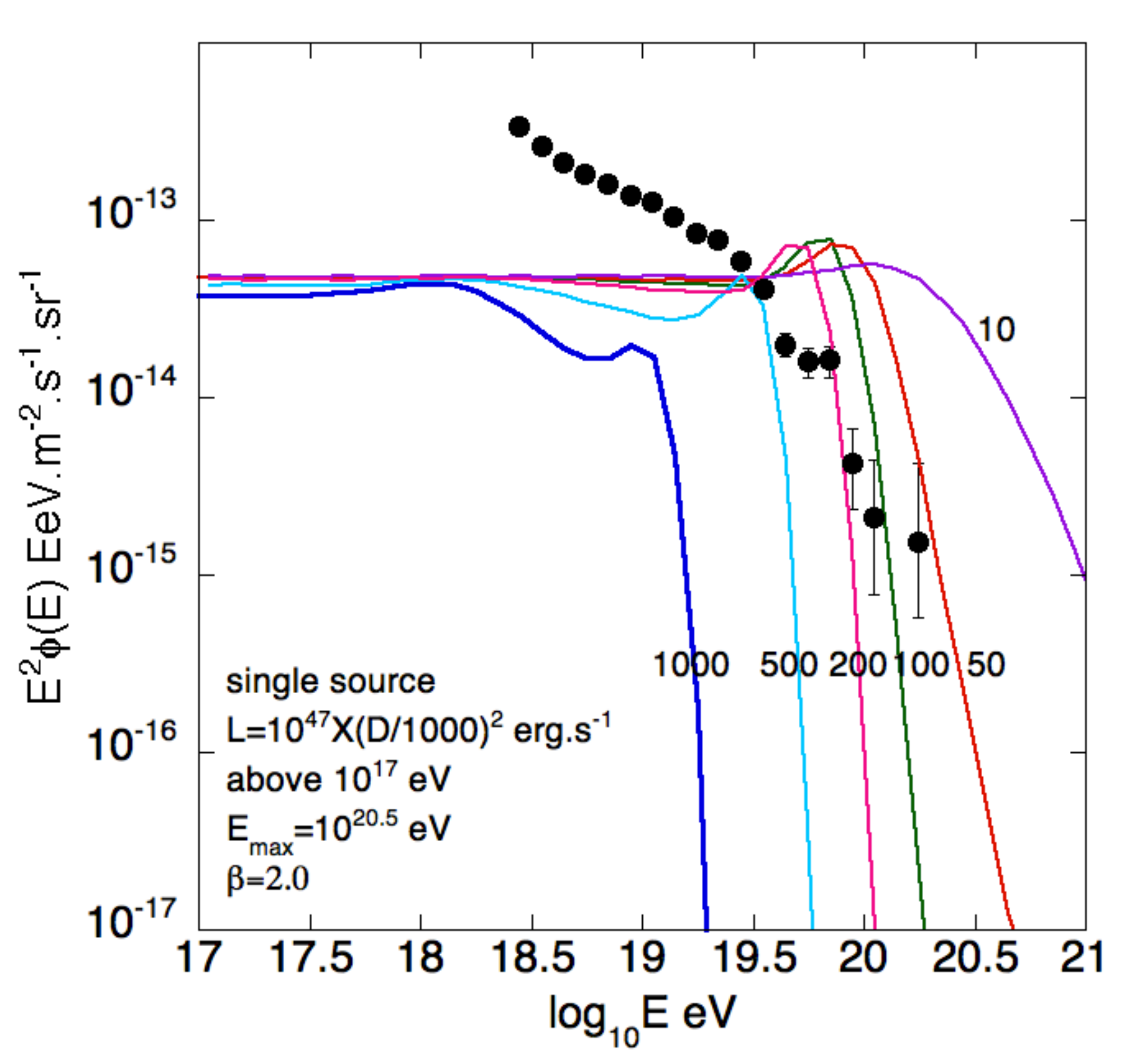}
%\vspace{-1.3cm}
\caption{Same as in Fig.~\ref{fig:source1000} assuming sources located at various distances D=5, 10, 20, 100, 200, 500, 1000 Mpc and luminosity of $\rm 10^{47}\times(D/1000Mpc)^2\,erg\,s^{-1}$. The top panel shows the photons and neutrino fluxes, the bottom panel the spectra compared with the Pierre Auger Observatory data. Only the pure proton case is considered.}
%\vspace{-0.6cm}
\label{fig:sourceneutdist}
\end{figure}

Photons in the GeV-TeV range for single individual sources of UHECRs have already been considered in Ferrigno et al. (2005), Armengaud et al. (2006) and more recently Essey et al. (2010) and Ahlers and Salvado (2010). In this energy range the cosmogenic photon flux displayed in the bottom panel of Fig.~\ref{fig:source1000} is very high and of course well within the range of existing and future gamma-ray observatories. But $\sim75\%$ of this flux are provided by the pair production mechanism, then most of the gamma-ray flux is produced far from the source (unless a very strong magnetic field is able to confine protons between $10^{18}$~eV and a few $10^{19}$~eV). The sole flux caused by pion production remains ample however and far above the gamma-ray observatories sensitivity for point-source detection. As pointed out in Gabici and Aharonian (2007) (see also Gabici, 2011) this flux is yet likely to be widely spread and isotropized as soon as the extragalactic field is not much lower than $10^{-12}$ G. For magnetic fields between $10^{-12}$ and $10^{-9}$ G (above this value, synchrotron losses influence the cascade development, see below, and Gabici and Aharonian ,2005, 2007) as discussed in Gabici and Aharonian (2007) and assuming an isotropic cosmic ray emission, an electromagnetic halo of typical radius $\sim 20$ Mpc, including most of the flux produced by the pion production mechanism, would form. It would spread the source image over an angular size on the order of $\sim$1 degree or more (depending on the precise value of the field, which is needed to estimate the 3D development of the electromagnetic cascade) for a source located at 1 Gpc. Then, the angular size of the source would exceed the typical point spread function of a few arc-minutes, which should be chosen for the future CTA gamma-ray experiment, but the high expected integrated flux might still be detectable. 

For greater values of the magnetic field, especially in the source environment, Gabici and Aharonian (2005) pointed out the possible detection of the synchrotron photons emitted during the first stages of the cascade development. This possibility was also recently studied in Kotera et al. (2011). In both studies, a source at 1 Gpc of luminosity $10^{46} \rm erg\,s^{-1}$ above $10^{19}$ eV is found to possibly yield detectable fluxes for Fermi or CTA (for instance, a flux of $~10^{-9}\rm GeV\,cm^{-2}\,s^{-1}$ spread over $\sim0.2$ deg). These synchrotron images are promising for the detection of distant UHECR accelerators and have the advantage that they are a signature of the pion production mechanism (see discussions in Gabici and Aharonian, 2005 and Kotera et al. 2011), unlike cascades in the case of distant sources (see below for the local universe). However, their signal is weaker in the TeV range than in the GeV, although their detection with CTA should still be possible. However, if the matter density is quite high in the magnetized environment (see the example of a galaxy cluster in Kotera et al., 2009), the synchrotron signal could be screened and overwhelmed by hadronic interactions of the confined low-energy cosmic rays.

Interestingly, GeV-TeV cascades were recently proposed  by Essey et al. (2010) (and were recently re-examined by Alhers and Salvado, 2011) as a possible interpretation of TeV signal observed by HESS of the AGN 1ES0229+200 (z=0.14). Assuming a point-source cascade signal, Essey et al. estimate the implied source luminosity in cosmic ray protons above $10^{16}$ eV to be between $\sim10^{46}$ and $10^{49}$ $\rm erg\,s^{-1}$ (depending on the maximum energy assumed). If confirmed, this interpretation would not be a signature of the acceleration of particles above the pion production threshold (because this signal could be produced by the sole pair production mechanism), but would involve either extremely weak magnetic fields ($\leq10^{-14}$ G) on very large scales or an even greater source luminosities (to allow the same flux to be within HESS point spread function).

One could argue that because the neutrino flux forms quickly, similar fluxes could be found by assuming weaker source luminosity $\rm L_D=10^{47}\times(D/1000)^2\,erg\,s^{-1}$. Neutrino and photon fluxes for sources located at distances D=5, 10, 20, 50,  100, 200, 500, 1000 Mpc assuming a luminosity $\rm L_D$ above $10^{17}$ eV, are shown in Fig.~\ref{fig:sourceneutdist}. For neutrinos only, fluxes for 10 and 1000 Mpc are shown for clarity. Neutrino fluxes become completely similar to those at 1 Gpc above $\sim20$ Mpc, slightly shifted to higher energies because of the lower redshifting of the neutrino and the higher energy threshold implied by the smaller distance of the source. These fluxes requiring lower luminosity could then be detected if neutrino observatories were able to reach a sensitivity around $10^{-8}$ $\rm erg\,s^{-1}$ above $10^{17}$ eV. However, as can be seen in the bottom panel of Fig.~\ref{fig:sourceneutdist}, unlike for a source at 1 Gpc, for which energy losses keep the propagated spectrum safely below the experimental data, a source located at D$\leq500$ Mpc would overshoot the cosmic ray spectrum for luminosities on the order of $L_D$\footnote{For the farthest distances it could be argued that the limited lifetime of the source (typically a few million or tens of million years for AGNs) and a hypothetical extragalactic magnetic field might cause a dilution of the flux during the propagation. This is quite unlikely for sources located at less than 100 Mpc, however, and our discussion cannot include bursting sources.}. For a reasonable contribution to the cosmic ray spectrum, say on the order of 10\% around $5\times10^{19}$ eV, one would have to assume a luminosity $\sim20$ times lower, \emph{e.g.} around $\rm \sim5\times10^{41}\,erg\,s^{-1}$ above $10^{17}$ eV (assuming $\beta=2.0$). Then, the flux would drop around $\rm 5\times10^{-10} \,GeV\,cm^{-2}\,s^{-1}$, which should be a lot more difficult to spot. Yet, reaching this sensitivity would make neutrino observatories extremely constraining for the potential accelerators in the local universe.  For the time being, the detection of an extremely powerful and distant source (which might be the prime candidate for proton acceleration above $10^{20}$ eV, see above) seems to be a more realistic goal. 

\begin{figure}[!t]
\includegraphics[width=1.0\columnwidth]{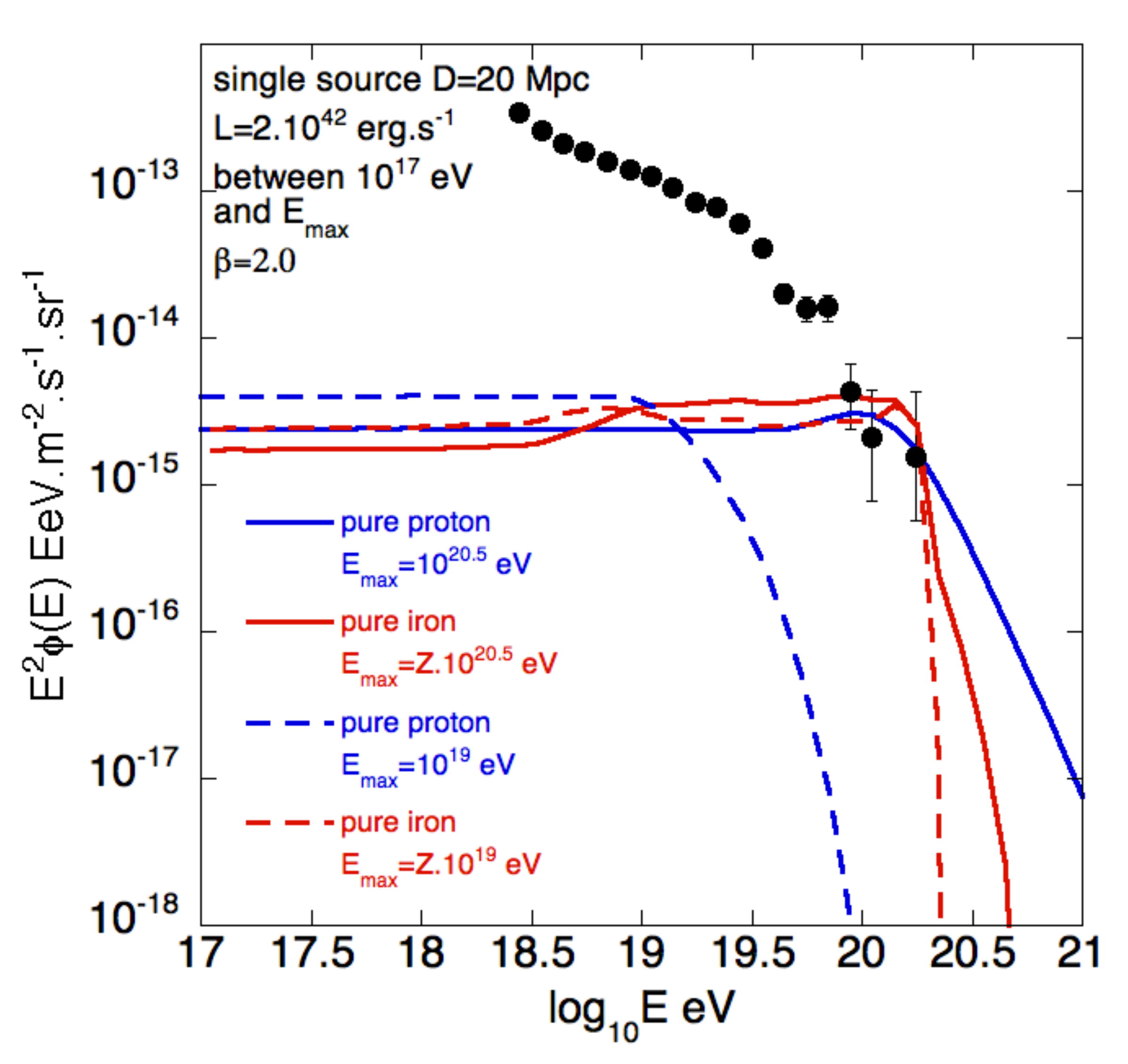}
\hfill
\includegraphics[width=1.0\columnwidth]{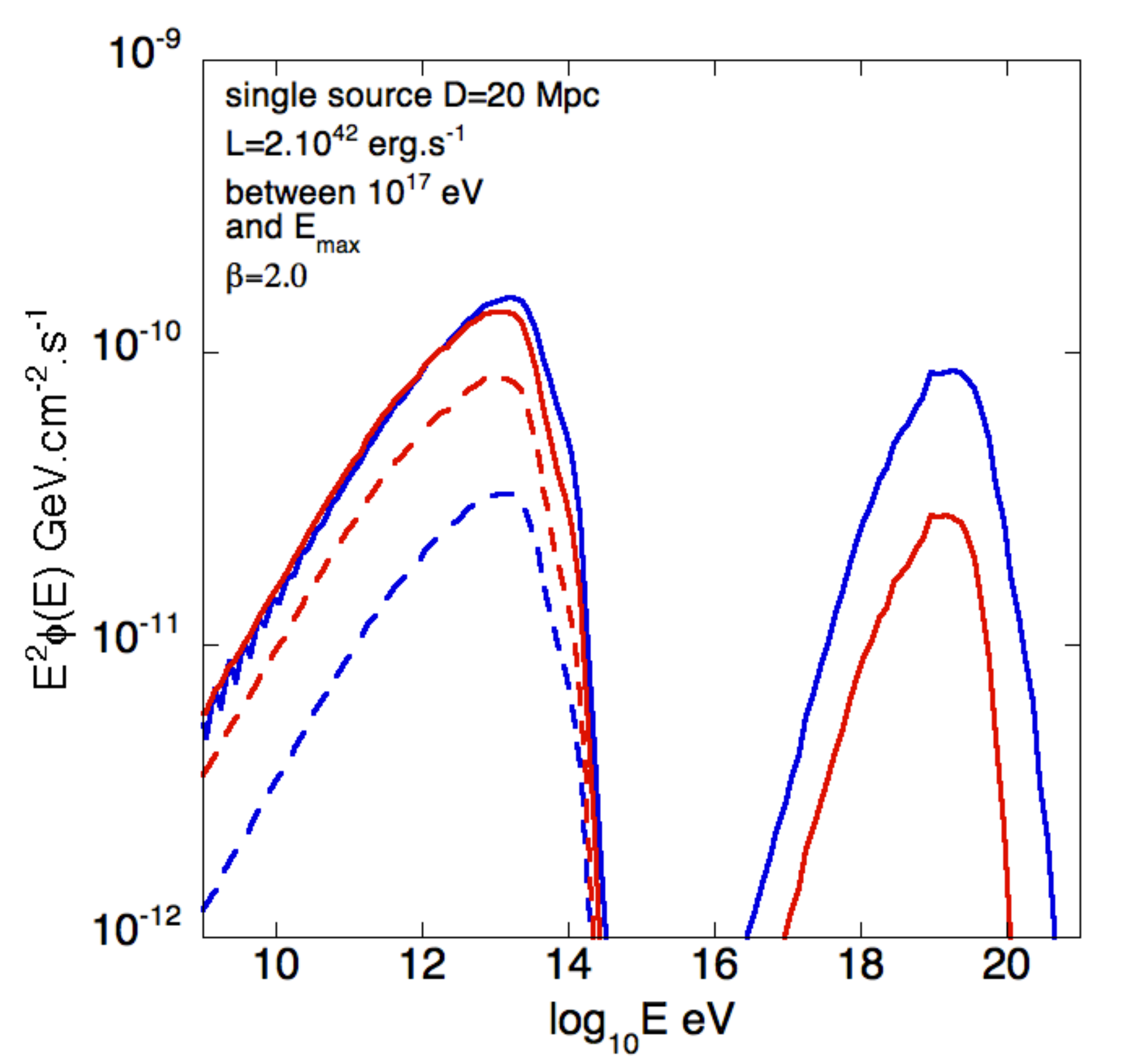}
%\vspace{-1.3cm}
\caption{Same as in Fig.~\ref{fig:source1000}, we now take a source at 20 Mpc with a luminosity 20 times lower than in Fig.~\ref{fig:sourceneutdist}. Pure proton and pure iron cases are considered as well as maximum energies of $\rm Z\times10^{20.5}$ eV and $\rm Z\times10^{19}$ eV.}
%\vspace{-0.6cm}
\label{fig:final}
\end{figure}

The corresponding cosmogenic photon fluxes (UHE and GeV-TeV) are also displayed in the top panel of Fig.~\ref{fig:sourceneutdist}. The GeV-TeV flux becomes lower as the source distance decreases. This is because the contribution of the pair production mechanism decreases. The higher contribution of the pion decay increases the average energy of secondaries that initiate the cascade and consequently the electromagnetic cascades take longer to develop down to the TeV range (magnetic field effects would of course make this discussion more complicated) which in turn makes the spectra harder, but the fluxes lower. As mentioned before, to obtain the sources that contribute to a level of $\sim10\%$ of the flux around $5\times10^{19}$ eV, one needs fluxes lower by a factor of $\sim20$. In the last case, they are likely to remain above the future CTA point source sensitivity. However, as mentioned above, (and following the discussion of Gabici and Aharonian, 2007), the effect of a potential extragalactic magnetic field on the angular size of the gamma-ray image of the source is expected to be much worse for nearby sources, making the detection of their signal quite hypothetical.

Finally, ignoring the extragalactic magnetic fields, we conclude with a discussion on the general impact of the composition and the maximum energy. Fig.~\ref{fig:final} shows the cosmic ray output (top panel) and the cosmogenic photon flux assuming a source located at 20 Mpc and a luminosity $\rm L=2\times10^{42}\,erg\,s^{-1}$ between $10^{17}$ eV and $\rm E_{max}$. We studied two source compositions, namely pure proton and pure iron, and two maximum energies, $\rm Z\times10^{20.5}$~eV and $\rm Z\times10^{19}$~eV. As mentioned above, with this luminosity the contribution to the total cosmic ray spectrum becomes plausible (between $\sim10$ and 20\% at $5\times10^{19}$~eV depending on the model). For the highest maximum energy the pure proton and pure iron fluxes are once again very close at low energies. For the low $\rm E_{max}$ sources ($\rm E_{max}=Z\times10^{19}$ eV, well below the pion production threshold, the proton flux drops by a factor of $\sim5$ compared to the high $\rm E_{max}$ case. The drop in the flux is, in this case, limited because the pair production mechanism, although less efficient for nearby sources, emits lower energy secondaries that cascade more rapidly to TeV energies, whereas the high-energy secondaries resulting from the pion production mechanism are more likely to remain at high energies if they are emitted close to the observer. In the low $\rm E_{max}$ scenario for the pure iron composition, the fluxes are even higher and closer to high $\rm E_{max}$ scenarios. Indeed, besides secondary protons, which are efficiently emitted by iron nuclei close to the maximum energy, the flux is produced quite efficiently by the pair production mechanism above $\sim6\,10^{19}$ eV with a loss length $\rm \sim Z^2/A$ times shorter than for protons at the same Lorentz factor (see Allard et al., 2006, for the contribution of the different energy loss processes on iron nuclei and discussion in Alhers and Salvado, 2011). The precise difference between the different compositions or maximum energies of course depends on the assumed spectral index, but these examples show that besides the probable angular spread that is likely to prevent the detection of these TeV images, GeV-TeV fluxes suffer to some extent from a degeneracy on physical parameters such as the  composition (which should also apply to neutrino sources) or maximum energy, the latter preventing in principle the use of these fluxes as a signature of sources accelerating cosmic rays above the pion production threshold, $\sim10^{20}$ eV per nucleon (unless a neutrino, UHE photon, of a UHE cosmic ray counterpart is also found). On the other hand,  a positive detection would bring unprecedented constraints on the extragalactic magnetic fields in the local universe and certainly allow a clear identification of the source.

%\begin{figure*}[t]
%\centering
%\hfill\includegraphics[height=6cm]{Dbas.pdf}\hfill
%\includegraphics[height=6cm]{Dhaut.pdf}\hfill~
%\caption{IEDC for protons of various energies, from $10^{16}$ to $10^{20}$ eV, in a 10 nG field with a maximum wavelength $\lambda_{\mathrm{max}}$=1 Mpc.}
%\label{IECD}
%\end{figure*}

\section{Conclusion}

We have considered the production of cosmogenic secondary photons and neutrinos during the extragalactic propagation of UHE protons and nuclei. We discussed the constraints obtained from {the Fermi} observations of the diffuse gamma-ray flux. We found that, because the possibility of a high contribution of UHECR to the diffuse gamma-ray flux is currently not ruled out, significant UHECR neutrino fluxes observable by present observatories such as IceCube or the Pierre Auger Observatory can still be expected and would be of prime interest for the field. The UHE cosmogenic photons could also be expected for optimistic astrophysical assumptions. These fluxes are not constrained by the experimental diffuse gamma-ray flux because they show little dependance on the evolution of the source luminosity evolution unlike UHE neutrinos. We also discussed the influence of the composition evolution at the highest energies. We pointed out that scenarios for which the maximum energy at the source is limited (well below $10^{20}$ eV per nucleon) are expected to have their UHECR neutrino flux strongly suppressed above $10^{17}$ eV. However, we found that the likely variation of the maximum from source to source could yield an observable neutrino flux, if rare powerful accelerators (e.g, source able to accelerate particles above $10^{20}$ eV per nucleon) show a strong cosmological evolution of their luminosity and contribute at the level of $\sim10\%$ to the cosmic ray flux around $10^{19}$~eV.

We also considered the possibility of observing cosmogenic secondaries from individual sources to constrain the origin of highest energy particles. Distant and very powerful sources were found to be prime candidates, and neutrino flux along with synchrotron images (Gabici and Aharonian, 2005) would be unambiguous signatures of the acceleration of particles above the threshold of pion production. These observations would not strongly constrain the source composition, though. But observations of sources in the local universe would require, with the current experimental capabilities, too high contributions of these individual sources to the UHE cosmic ray spectrum. This can be alleviated, however, with improved sensitivities (it would require point source sensitivities below $\rm10^{-9}\, GeV\,cm^{-2}\,s^{-1}$ above $10^{17}$ eV for neutrino observatories, for instance), which would at the same time decrease the luminosity requirements for distant sources.  GeV-TeV photons from electromagnetic cascade, taking advantage of the upcoming very sensitive CTA observatory, could be detected even in the local universe by assuming a straight line propagation of the electromagnetic cascades. However, their detectability strongly depends on the strength of the extragalactic magnetic field and sources in the local universe are likely to become very hard to detect if the latter exceeds $\sim10^{-12}$ G in the local universe. For a positive detection strong constraints on the extragalactic magnetic field could of course be obtained as well as a precise location of the source -- a critical milestone in high-energy astrophysics. However, these cascades are not a signature of the acceleration of cosmic rays above the threshold of pion production or of the source composition. Consequently, counterparts (primarily from cosmic ray observatories) would be needed to fully understand the signal.

We conclude that the observation of cosmogenic secondaries either by their diffuse flux or individual sources can be expected from current or next-generation instruments. Importantly, they would represent invaluable additions to observations of the UHECR sky made by the Pierre Auger Observatory, the Telescope Array and the future JEM-EUSO in their quest for solving the long-standing UHECR problem and identifying its astrophysical source.


\begin{thebibliography}{}

\bibitem{HiRes} Abbasi, R. U. et al. [HiRes collaboration] 2004, Phys. Rev. Let., 92, 151101

\bibitem{Abbasi} Abbasi R. U. et al. [IceCube collaboration] 2009, Phys. Rev. Let., 103, 221102

\bibitem{Abbasi11a} Abbasi R. U. et al. [IceCube collaboration] 2011, Phys. Rev. D, 83(11), 092003

\bibitem{FermiIR} Abdo A. A. et al. [Fermi Collaboration] 2010, ApJ, 723, 1082

\bibitem{FermiGRBack10}Abdo A. A. et al. [Fermi Collaboration] 2010, Phys. Rev. Let, 104, 101101

\bibitem{Fermidiff}Abdo A. A. et al. [Fermi Collaboration] 2010, ApJ, 720, 435

\bibitem{Augerer2004}Abraham et al. [Pierre Auger Collaboration] 2004,  Nuclear Instruments and Methods in Physics Research Section A, 523, 50

\bibitem{Augerer2008}Abraham et al. [Pierre Auger Collaboration] 2008,  Astroparticle Physics, 29, 243

\bibitem{Augerphot2009} Abraham J. et al. [Pierre Auger Collaboration] 2009, Astroparticle Physics, 31, 399Ð406

\bibitem{Augerneut2009} Abraham J. et al. [Pierre Auger Collaboration] 2009, Phys. Rev. D, 79(10), 102001

\bibitem{Augerer2010}Abraham et al. [Pierre Auger Collaboration] 2010,  Phys. Rev. Let., 104, 91101

\bibitem{Augersp2010}Abraham et al. [Pierre Auger Collaboration] 2010, Physics Letters B, 685, 239

\bibitem{AugerICRC2011} Abraham et al.  [Pierre Auger Collaboration] July 2011, astro-ph.HE, Contributions to the 32nd International Cosmic Ray Conference, Beijing, China, August 2011.

\bibitem{AugerNeut2011}Abreu P. et al. [Pierre Auger Collaboration] 2011, proc. of the $\rm32^{nd}$ international cosmic-ray conference, Beijing (China), arXiv:1107.4805v1.

\bibitem{Achterberg2002} Achterberg A., 2002, proceedings of les Houches summer school, Accretion disks, jets and high-energy phenomena in astrophysics.

\bibitem{Ahlers2009} Ahlers M., Anchordoqui L. A. and Sarkar S. 2009, Phys. Rev. D 79(8), 083009

\bibitem{Ahlers2010} Ahlers M., Anchordoqui L. A., Gonzalez-Garcia M. C., Halzen F. and Sarkar S. 2010, Astroparticle Physics, 34, 106

\bibitem{Ahlers2011} Ahlers M., Salvado J., 2011, [arXiv:1105.5113 [astro-ph.HE]]

\bibitem{Allard05a} Allard D., Parizot E., Khan E., Goriely S. and  Olinto A.~ V. 2005, A\&A Letters,  443, 29

\bibitem{DenisNeut2006} Allard D., Ave M., Busca N., Malkan M. A., Olinto A. V., Parizot E., Stecker F. W. and
Yamamoto T. 2006, Journal of Cosmology and Astro-Particle Physics 9, 5

\bibitem{Denis2007} Allard D., Parizot E. and Olinto A. V. 2007, Astroparticle Physics 27, 61Ð75

\bibitem{Denis2008}Allard D., Busca N. G., Decerprit G., Olinto A. V. and Parizot E. 2008, Journal of Cosmology and Astro-Particle Physics, 10, 33

\bibitem{Denis2009}Allard D. 2009, [arXiv0906.3156]

\bibitem{Allard2009}Allard D. and Protheroe R. J. 2009, A\&A, 502, 803

\bibitem{Anchordoqui2007} Anchordoqui L. A., Goldberg H., Hooper D., Sarkar S., Taylor A. 2007, Phys. Rev. D, 76, 123008

\bibitem{Angelov} Anguelov V., Petrov S., Gurdev L., and Kourtev J. 1999, J. Phys., G25:1733

\bibitem{Arons} Arons J. 2003, ApJ, 589, 871

\bibitem{Ave2005} Ave M., Busca N., Olinto A. V., Watson A. A. and Yamamoto T. 2005, Astroparticle Physics, 23, 19Ð29

\bibitem{BereOriginal} Berezinsky  V.S. and  Zatsepin, G.T. 1969, Phys. Lett. B, 28, 423

\bibitem{Bere1975} Berezinsky V. S. and Smirnov A. 1975, Astrophysics and Space Science, 32, 461

\bibitem{BerezinskyDip} Berezinsky V. S., Gazizov A. and Grigorieva S. 2006 Phys. Rev. D, 74(4), 043005

\bibitem{BereNeut2009} Berezinsky V. S. 2009, Proc. 4th Int. Workshop "Neutrino Oscillations in Venice", ed. Milla Baldo Ceolin, 137-158, [arXiv:0901.1428]

\bibitem{Berezinsky:2010xa}  Berezinsky V., Gazizov A., Kachelriess M. and Ostapchenko~S. 2010, Phys.\ Lett.\  B, 695, 13
  
\bibitem{BereFermi2010} Berezinsky V. S., Gazizov A., Kachelriess M., Ostapchenko S. 2010, Phys. Lett. B, 695, 13

\bibitem{Clark1970} Clark T. A.,  Brown L. W., and Alexander J. K. 1970, Nature, 228, 847

\bibitem{Tanguy} d'Enterria D., Engel R., Pierog T., Ostapchenko S., Werner K. 2011, Astropart.\ Phys.\, 35, 98-113

\bibitem{Engel2001} Engel R., Seckel D. and Stanev T. 2001, Phys. Rev. D, 64(9), 093010

\bibitem{Epstein}Blasi P., Epstein R.~I. and  Olinto A.~V. 1999, ApJ, 533, L123

\bibitem{Essey10}Essey W., Kalashev O. E., Kusenko A., Beacom J. F. 2010, Phys. Rev. Let., 104, 141102

\bibitem{Ferrigno2005}Ferrigno C., Blasi P. and  de Marco D. 2005, Astroparticle Physics, 23, 211

\bibitem{Gabici2005}Gabici S. and Aharonian F. 2005, Phys. Rev. Let., 95, 251102

\bibitem{Gabici2007}Gabici S. and Aharonian F. 2007, Astrophysics and Space Science, 309, 465

\bibitem{Gabici2011}Gabici S. 2011, Gamma ray astronomy and the origin of galactic cosmic rays, habilitation \`a diriger des recherches de l'universit\'e Paris VII.

\bibitem{Gelmini2005} Gelmini G., Kalashev O.,  Semikoz D. 2008, J.\ Exp.\ Theor.\ Phys.\, 106, 1061-1082

\bibitem{Gelmini2007a} Gelmini G., Kalashev O.,  Semikoz D. 2007, Journal of Cosmology and Astroparticle Physics, 11, 002

\bibitem{Gelmini2007b} Gelmini G., Kalashev O.,  Semikoz D. 2007, Astroparticle Physics, 28, 390

\bibitem{G66} Greisen K. 1966, Phys. Rev. Lett., 16, 748

\bibitem{Haug}Haug E. 1975, Zeitschrift Natur-forschung Teil A, 30, 1546

\bibitem{Hill85} Hill C. T. and Schramm D. N. 1985, Phys. Rev. D, 31, 564

\bibitem{Hillas} Hillas A. M. 1984, Annual Review of Astronomy and Astrophysics, 22, 425
      
\bibitem{Hooper2005}Hooper D., Taylor A. and Sarkar S. 2005, Astroparticle Physics, 23, 11Ð17

\bibitem{Hooper2010}Hooper D., Taylor A. and Sarkar S. 2010, Astroparticle Physics, 34, 340

\bibitem{Hopkins} Hopkins A. M. and Beacom J. F. 2006, ApJ, 651, 142Ð154

\bibitem{Jones} Jones F. C. 1968, Phys. Rev., 167(5), 1159

\bibitem{Kalashev2002}Kalashev O. E., Kuzmin V. A., Semikoz D., Sigl G. 2002, Phys. Rev. D, 66, 63004

\bibitem{Kalashev2007}Kalashev O., Semikoz D.,  Sigl G. 2007, Phys. Rev. D., 79, 63005

\bibitem{KS2006} Kachelriess M. and Semikoz D. V 2006, Physics Letters B, 634, 143Ð147

\bibitem{Kelner}Kelner S. R. and Aharonian F. A. 2008, Phys. Rev. D, 78, 034013

\bibitem{Khan2005}Khan E., Goriely S.,  Allard D., Parizot E.,  Suomijarvi T.,  Koning A. J., Hilaire S.,  Duijvestijn M. C. 2005, Astroparticle Physics, 23, 191

\bibitem{Kneiske2004} Kneiske T. M., Bretz T., Mannheim K. and Hartmann D. H. 2004, A\&A, 413, 807Ð815

\bibitem{Kotera2010}Kotera K.,  Allard D. and Olinto A. V. 2010, Journal of Cosmology and Astroparticle Physics, 10, 013

\bibitem{Kotera2009}Kotera K.,  Allard D.,  Murase K.,  Aoi J. , Dubois Y.,  Pierog T.,  Nagataki S. 2009, ApJ, 707, 370

\bibitem{Kotera2011}Kotera K.,  Allard D. and Lemoine M. 2011,  A\&A, 527, 54

\bibitem{kuempel2009}Kuempel D., Kampert K. H. and Risse M. 2009, proceedings of the $\rm31^{st}$ ICRC, Lodz (Poland), [arXiv:0906.3099]

\bibitem{Lee1998}Lee S. 1998, Phys. Rev. D., 58, 43004

\bibitem{Lemoine09}Lemoine M. and Waxman E. 2009, Journal of Cosmology and Astroparticle Physics, 11, 009

\bibitem{Mastichiadis} Mastichiadis A. 1991, Nov. 15, Royal Astronomical Society Monthly Notices, 253, 235-244

\bibitem{Medina} Medina-Tanco G. et al. [JEM-EUSO collaboration] 2009, [arXiv:0909.3766]

\bibitem{Mucke}Mucke A., Engel R., Rachen J. P., Protheroe R. J., and Stanev T. 2000, Comp. Phys. Com., 124, 290

\bibitem{Nagano92}Nagano N. et al. 1992, Nucl. Part. Phys., 18, 423

\bibitem{QFT}  Peskin M. E.  and Schroeder D. V. 1995, Addison-Wesley, An Introduction to Quantum Field Theory

\bibitem{Protheroe1996}Protheroe R. J and  Johnson P. A. 1996, Astroparticle Physics, 4, 253

\bibitem{Rachen96} Rachen J. P. 1996, Interaction processes and statistical properties of the propagation of cosmic rays in photon backgrounds, Ph.D. Thesis, Bonn University

\bibitem{Risse:2007sd}Risse M. and Homola P. 2007, Mod. Phys. Lett. A,  22, 749

\bibitem{Seckel2005} Seckel D. and Stanev T. 2005, Physical Review Letters, 95(14), 141101

\bibitem{Sigl1999}Sigl G.,  Lee S.,  Bhattacharjee P.,  Yoshida S. 1999, Phys. Rev. D, 59, 043504

\bibitem{Strong1973} Strong A. W. et al. 1973, Nature, 241, 109

\bibitem{Strong1974} Strong A. W. et al. 1974, J. Phys. A: Math. Nucl. Gen. 7 120

\bibitem{EGRET1} Sreekumar P. et al. 1998, ApJ, 494, 523

\bibitem{Stanev2005} Stanev T., de Marco D., Malkan M. A. and Stecker F. W. 2006, Phys. Rev. D, 73(4), 043003

\bibitem{Stecker1979}Stecker F. W. 1979, ApJ, 228, 919Ð927

\bibitem{Stecker2006}Stecker F. W., Malkan M. A. and Scully S. T. 2006, ApJ, 648, 774Ð783

\bibitem{EGRET2} Strong A. W., Moskalenko I. V. and Reimer O. 2004, ApJ 613, 956

\bibitem{Takami2009}Takami H, Murase K, Nagataki S and Sato K 2009, Astroparticle Physics, 31, 201Ð211

\bibitem{Taylor2009}Taylor A. M. and Aharonian F. A. 2009, Phys. Rev. D, 79(8), 083010

\bibitem{Taylor2010}Taylor A. M.,  Hinton J. A. ,  Blasi P. and Ave M. 2009, Phys. Rev. Let., 103, 51102

\bibitem{AugerICRC2009}Tiffenberg J., [Pierre Auger Collaboration] 2009, proceedings of the $\rm31^{st}$ ICRC Lodz (Poland), [arXiv0906.2347]

\bibitem{Wall}Wall J. V., Jackson C. A., Shaver P. A., Hook I. M. and Kellermann K. I. 2005, A\&A, 434, 133-148

\bibitem{Wang2011}Wang X., Liu R. and Aharonian F. 2011, ApJ , 736, 112

\bibitem{WW}Wibig T. and Wolfendale A. W. 2005, J.\ Phys.\ G, 31, 255

\bibitem{Yoshida1993} Yoshida S. and Teshima M. 1993,  Prog. Theor. Phys., 89, 833

\bibitem{ZK66} Zatsepin G. T. and Kuzmin V. A. 1966,  Sov. Phys. JETP Lett., 4, 78

\bibitem{Zdziarsky} Zdziarski A. A. and Svensson R. 1989, Nucl. Phys. Proc. Suppl., 10B:81Ð88, 1989

\end{thebibliography}
\end{document}